\documentclass[aip,amsmath,nofootinbib,amssymb,reprint]{revtex4-1}
\usepackage{graphicx}
\usepackage{dcolumn}
\usepackage{bm}
\usepackage[utf8]{inputenc}
\usepackage[T1]{fontenc}
\usepackage{mathptmx}
\usepackage{etoolbox}
\usepackage{color}

\draft 

\makeatletter
\def\@email#1#2{%
 \endgroup
 \patchcmd{\titleblock@produce}
  {\frontmatter@RRAPformat}
  {\frontmatter@RRAPformat{\produce@RRAP{*#1\href{mailto:#2}{#2}}}\frontmatter@RRAPformat}
  {}{}
}%
\makeatother

\begin{document}

\preprint{AIP/123-QED}


\title[SPH Physically Reconsidered - The Relation to Explicit LES and the Issue of Particle Duality]{Smoothed Particle Hydrodynamics Physically Reconsidered - The Relation to Explicit Large Eddy Simulation and the Issue of Particle Duality}
\author{M. Okraschevski}
    \email{max.okraschevski@kit.edu}
    \affiliation{Institute of Thermal Turbomachinery, Karlsruhe Institute of Technology, Kaiserstraße 12, 76131 Karlsruhe, Germany}
\author{N. Buerkle}%
    \affiliation{Institute of Thermal Turbomachinery, Karlsruhe Institute of Technology, Kaiserstraße 12, 76131 Karlsruhe, Germany}
\author{R. Koch}%
    \affiliation{Institute of Thermal Turbomachinery, Karlsruhe Institute of Technology, Kaiserstraße 12, 76131 Karlsruhe, Germany}%
\author{H.-J. Bauer}
    \affiliation{Institute of Thermal Turbomachinery, Karlsruhe Institute of Technology, Kaiserstraße 12, 76131 Karlsruhe, Germany}%

\date{\today}

\begin{abstract}
In this work we will identify a novel relation between Smoothed Particle Hydrodynamics (SPH) and explicit Large Eddy Simulation (LES) using a coarse-graining method from Non-Equilibrium Molecular Dynamics (NEMD). While the current literature points at the conclusion that characteristic SPH issues become restrictive for subsonic turbulent flows, we see the potential to mitigate these SPH issues by explicit subfilter stress (SFS) modelling. We verify our theory by various simulations of homogeneous, isotropic turbulence (HIT) at $Re=10^4$ and compare the results to a Direct Numerical Simulation (DNS) reported in Ref. \cite{Dairay_2017} Although the simulations substantiate our theory, we see another issue arising, which is conceptually rooted in the particle itself, termed as \emph{Particle Duality}. Finally, we conclude our work by acknowledging SPH as coarse-graining method for turbulent flows, highlighting its capabilities and limitations.
\end{abstract}


\maketitle 

\section{\label{sec:Introduction}Introduction}
Since its first introduction for astrophysical flow problems by Lucy and Gingold \& Monoghan in 1977 \cite{Lucy_1977, Gingold_1977}, the success of SPH as a viable Lagrangian method in the Computational Fluid Dynamics (CFD) community is undeniable. In the last decades there was a considerable research effort to increase the fundamental maturity of the method, summarized in several reviews \cite{Monoghan_2005, Springel_2010, Price_2012, Ye_2019, Lind_2020, Sigalotti_2021}, which was in parallel accompanied by progress regarding applications of higher complexity, e.g. Ref. \cite{Monoghan_2012, Shadloo_2016, Chaussonnet_2020}.

One of the most fundamental problems of classical Lagrangian SPH is that it suffers from zeroth order errors, which result in a substantial amount of noise compared to grid based Eulerian methods \cite{Bauer_2012, Hopkins_2015}. Physically, this noise causes excessive dissipation \cite{Ellero_2010, Bauer_2012, Colagrossi_2013, Hopkins_2015}  by numerically induced small scale vorticity \cite{Colagrossi_2013}. Although it could already be hypothesized based on the work of Ellero \emph{et al.} \cite{Ellero_2010} that this might become a severe issue for subsonic turbulence, a rigorous and detailed analysis proving this fact for forced homogeneous, isotropic turbulence (HIT) was presented in the seminal work of Bauer \& Springel \cite{Bauer_2012}. Up to date, the shortcomings of SPH for subsonic turbulence as discussed by the authors persist, namely that large scale turbulent structures can be qualitatively captured but at comparably high computational cost taking alternative CFD methods into account. This is quite unsatisfactory given that turbulence is a key aspect of most fluid flows. 

However, it might be argued that the results obtained by Bauer \& Springel \cite{Bauer_2012} are the consequence of a missing turbulence model and that they are only valid for underresolved Direct Numerical Simulations (uDNS). There were several publications on turbulence modelling in SPH \cite{Dalrymple_2006, Violeau_2007, Monoghan_2011, Leroy_2014, Mayrhofer_2015, DiMascio_2017, Antuono_2021}, but most of them either show a marginal improvement or are rather inconclusive for three dimensional subsonic turbulence. The latter can be attributed to the fact that the models are tested only with scarce validation runs, on setups which contain complex boundaries adding other SPH specific uncertainties on top of the actual turbulent flow \cite{Vacondio_2021} or are validated for two dimensional turbulence, which behaves qualitatively different \cite{Bofetta_2012} and where the use of usual turbulence closure models is unjustified, e.g. Ref. \cite{Sukoriansky_1996, Awad_2009}. In our opinion, the most promising approach so far was presented in the pioneering work of Di Mascio \emph{et al.} \cite{DiMascio_2017} and only recently extended by Antuono \emph{et al.} \cite{Antuono_2021}. In these works the authors explore SPH from a Large Eddy Simulation (LES) perspective, which represents a natural option as already noticed three decades ago \cite{Bicknell_1991}. Despite the fact that the derived SPH-LES approach with its various additional terms is an important step for SPH towards turbulent flows, both works do not evaluate the foundations of the classical LES subfilter stress (SFS) in a SPH framework, which is the central quantity in LES. In our opinion, the latter objective is vital, because the classical Lagrangian SPH features a turbulent kinetic energy deficit \cite{Bauer_2012}, which questions the intention of introducing mostly dissipative SFS models from the beginning \cite{Rennehen_2021}.

The work presented in this paper focuses on connecting explicit LES and SPH with a coarse-graining method from Non-Equilibrium Molecular Dynamics (NEMD). Hence, this study can be viewed as a sequel of our recent publications \cite{Okraschevski_2021_1, Okraschevski_2021_2}. We will demonstrate that the only additional term emerging from our theory is the SFS term as it is known from Eulerian LES methods. This is contrary to the statement of Di Mascio \emph{et al.}, concluding that a proper LES interpretation of SPH necessitates the consideration of additional SPH exclusive terms, though the authors ascertain that these terms play a minor role \cite{DiMascio_2017, Antuono_2021}. Consequently, we will be able to discuss the rationality of SFS models for the SPH simulation of subsonic turbulent flows. We are not considering further heuristic noise-mitigating techniques. Most importantly, our work is motivated by the following central question: \emph{Can resolved large scale structures profit from the reduction of SPH typical small scale noise by explicit use of SFS models}?

In order to elaborate this hypothesis, this paper is structured as follows: We start with a short review of the main characteristics of subsonic HIT and how the large scale dynamics of such turbulent flows can entirely be described by coarse-graining regularization of the fluid dynamic balance equations \cite{Eyink_2018}. This technique is most commonly known as LES. Then, we will relate this coarse-grained picture of subsonic turbulence to SPH. We constitute that SPH can be viewed as a Lagrangian quadrature technique for the governing equations of explicit LES and discuss the significant implications of this approach. To verify our theory, we will subsequently present various results of subsonic HIT simulations at $Re=10^4$ and compare the results to a DNS solution reported in Ref. \cite{Dairay_2017}. Finally, we will draw a conclusion on the rationality of SFS models in SPH.

\section{\label{sec:Turbulence}LES: Coarse-grained Dynamics of Subsonic Turbulent Flows}

\begin{figure}[ht]
\includegraphics[width=3.3in]{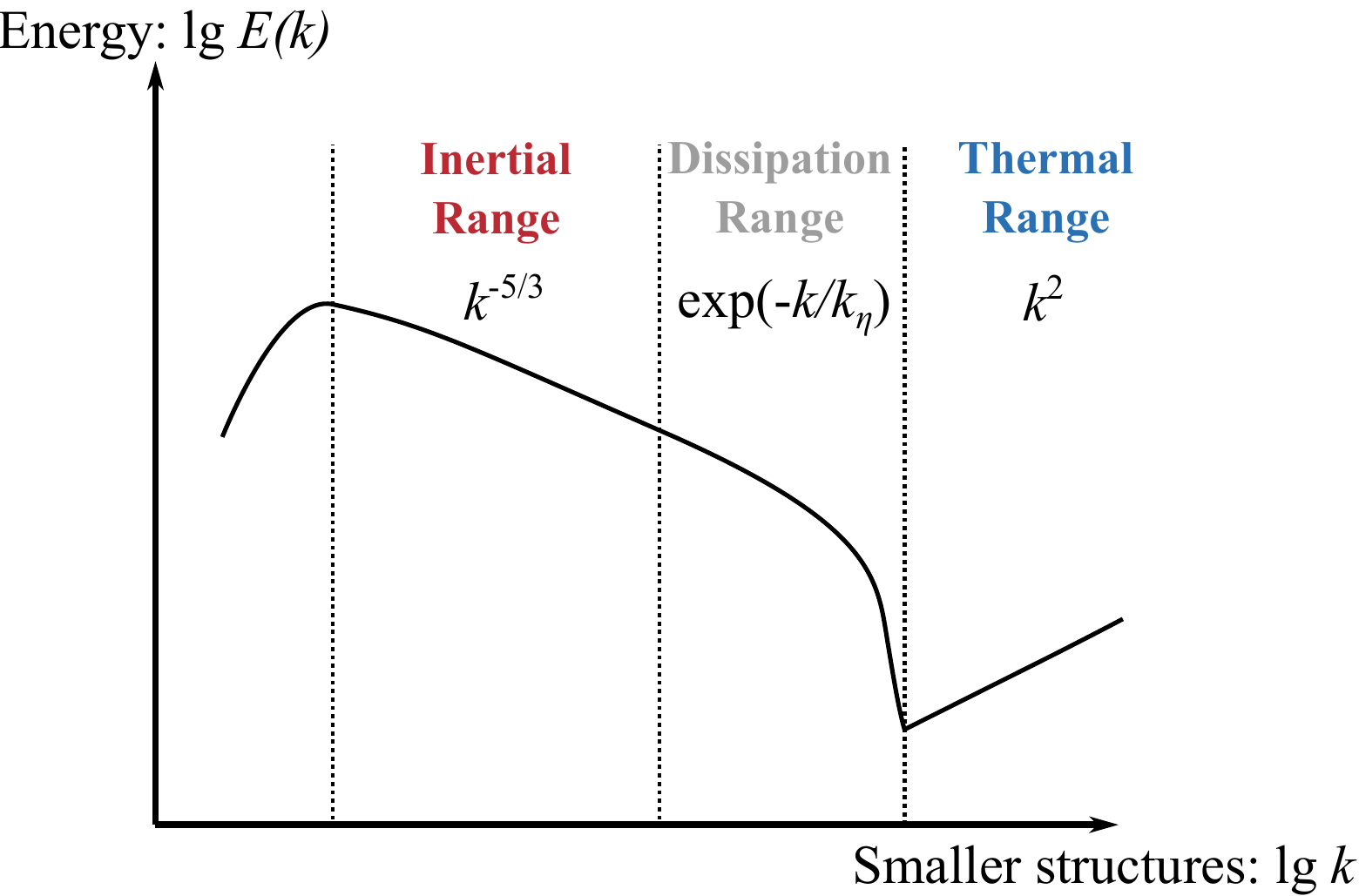}
\caption{Schematic of the turbulent kinetic energy spectrum for fully developed HIT in the strongly subsonic regime, including the thermal range beyond the classic dissipation range according to Ref. \cite{Bandak_2022}. }
\label{fig:00_SchematicSpectra}
\end{figure}

Despite the omnipresence and extraordinary beauty of turbulent flows, a comprehensive theory is still missing. However, it is agreed by the fluid dynamics community that the concept of the energy cascade is an important cornerstone of turbulence theory \cite{Eyink_2018}. The cascade process was metaphorically described by Richardson \cite{Richardson_1922} in 1922 for the first time, before it was quantified for incompressible HIT by Kolmogorov \cite{Kolmogorov_1941}, Obukhov \cite{Obukhov_1941}, Onsager \cite{Onsager_1945} and Heisenberg \cite{Heisenberg_1945} about 20 years later. For fully developed turbulence it was already demonstrated back then that, in a statistically averaged sense, a range of large scales exists in which the kinetic energy of velocity fluctuations of a specific wavenumber, namely $E(k)$, is transferred from larger to smaller scales in the absence of viscous dissipation effects. This range is known as inertial range \cite{Eyink_2018} and its scaling characteristics of 
\begin{equation}
    E(k) \sim k^{-5/3}
    \label{eq:HIT_Energy_Scaling}
\end{equation}
serves as an important benchmark for CFD solvers to prove their capability to reproduce large scale dynamics of strongly subsonic turbulent flows, e.g. Refs. \cite{Dairay_2017, Bauer_2012}.
In contrast, latest research activities on strongly subsonic HIT focus on the smallest scales of the cascade process, namely the dissipation range and beyond \cite{Smith_2015, Gallis_2021, Sreenivasan_2021, Bandak_2022, Gallis_2022, Bell_2022}. Among others, it has been rigorously argued by Sreenivasan \& Yakhot \cite{Sreenivasan_2021}, with the aid of a novel anomalous scaling theory, that the actual smallest turbulent length scale with wavenumber $k_\infty$ falls below the Kolmogorov scale with $k_\eta$. Another example is the importance of thermal fluctuations beyond the classical dissipation range, i.e. $k_\eta < k <  k_{mfp}$ with $k_{mfp}$ denoting the wavenumber of the mean free path, eventually leading to a $E(k) \sim k^{2}$ scaling \cite{Bandak_2022, Gallis_2022, Bell_2022}. Overall, all these insights can be vividly condensed in the turbulent kinetic energy spectrum as depicted in FIG. \ref{fig:00_SchematicSpectra}, highlighting the different turbulent regimes.

Although all scales are of significant importance for a holistic view of HIT, based on the work of Bauer \& Springel \cite{Bauer_2012} we can already conclude that classical Lagrangian SPH struggles to directly resolve turbulent small scale dynamics. Hence, it seems more convenient to combine SPH with a coarse-grained model, which intrinsically focuses on the turbulent large scales with its inertial range characteristics (Eq. (\ref{eq:HIT_Energy_Scaling})). This approach is more likely to be compatible, as the feedback of the smallest scales only has to be modelled and not resolved. A coarse-grained model of that kind can be derived by means of a technique from the Non-Equilibrium Molecular Dynamics (NEMD) community, \textcolor{black}{as we demonstrated in Ref. \cite{Okraschevski_2021_2}. In the following we will shortly summarize the main ideas of this generalization of Hardy's theory from 1982 \cite{Hardy_1982}.}

\begin{figure*}[t]
\includegraphics[width=5in]{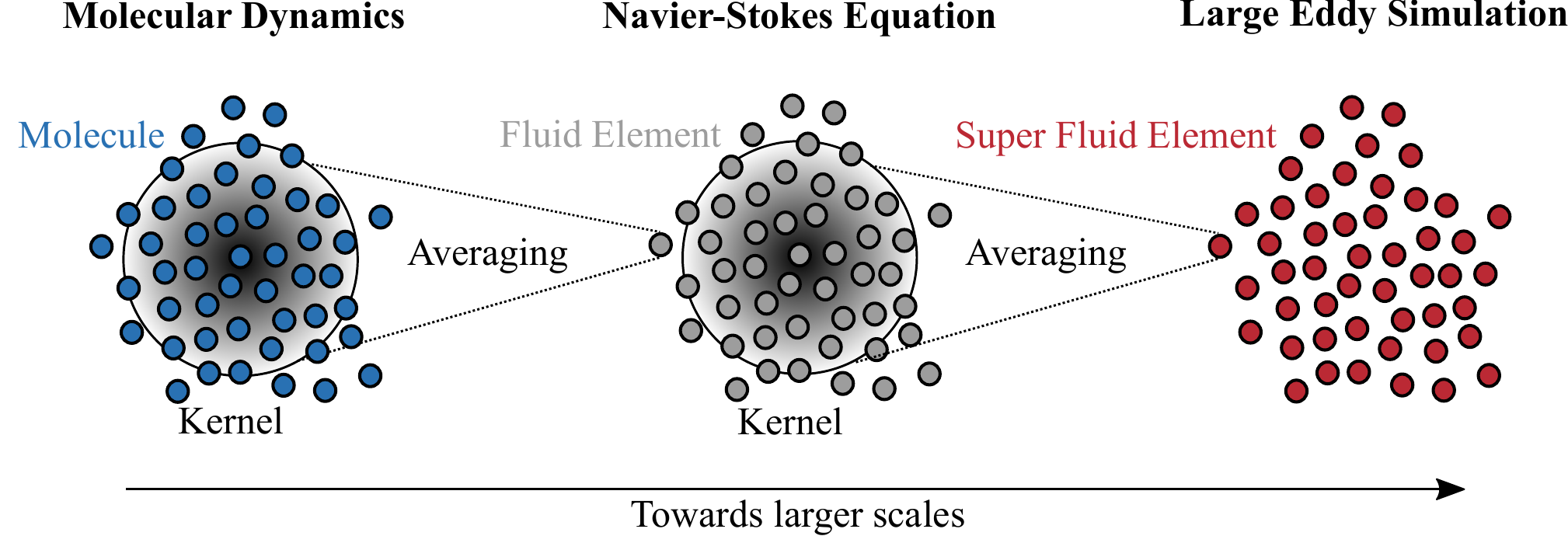}
\caption{Schematic of consecutive application of Hardy theory to different particle groups. Using a spherical kernel with size $\sim h$, smaller particles can be transferred into coarse-grained particles by appropriate averaging. As a consequence, axiomatic conservation properties, e.g. mass, momentum, energy, are scale-independently satisfied.}
\label{fig:01_HardyLES}
\end{figure*}

\textcolor{black}{The key aspect of the Hardy theory} is that it transfers arbitrary Lagrangian particles into coarse-grained particles by appropriate averaging, satisfying axiomatic conservation properties scale independently. The averaging is mathematically accomplished by the introduction of a normalized, symmetrical, positive, and monotonously decaying function $W_h$ with compact support $\textit{supp} \{ W_h \} \subset \mathbb{R}^3$, in short kernel. It is assumed that the latter is spherical and its spatial extent is quantified by the scalar index $h \in \mathbb{R}^+$.
Originally, as depicted in FIG. \ref{fig:01_HardyLES}, the Hardy theory was used to link the dynamics of discrete molecules and individual fluid elements governed by their continuum balance equations, e.g. Navier-Stokes for Newtonian flows. Even more important for this work is the fact that the Hardy theory can be generalized to a continuous set of Lagrangian particles as well, i.e. fluid elements (FIG. 2). This generalization leads to the governing equations of coarse-grained super fluid elements, which for $h=const$ are completely equivalent to the governing equations of LES. The latter represent a deterministic fluid flow model, which inherently focuses on scales above the kernel size and is able to capture the turbulence cascade \cite{Eyink_2018}. Thus, we interpret it as a physical description, which matches well with the SPH image for subsonic turbulence provided by Bauer \& Springel \cite{Bauer_2012}.

Although this LES perspective based on Lagrangian particles requires an additional mathematical effort compared to the common commutable filtering operation  \cite{Sagaut_2006}, it reveals a striking similarity between LES and SPH. A relation between the latter was already noticed 30 years ago by Bicknell \cite{Bicknell_1991}, but we take the view that our perspective on LES vividly strengthens this point. As common feature, both methods perform a coarse-graining of Lagrangian particles by means of a kernel and we will use this property in Sec. \ref{sec:SPH} to argue that SPH can be reinterpreted as a Lagrangian quadrature technique of the governing equations of LES. 

However, before we proceed with this objective, we will first concentrate on a fluid flow model, which is in general capable to describe isothermal, strongly subsonic HIT. Therefore, we select a Newtonian, barotropic fluid flow completely specified by its density, pressure and velocity field, namely $\rho, p$ and  $\mathbf{v}$, and apply the Hardy theory to it as detailed in Ref. \cite{Okraschevski_2021_2}. With $\mathbf{x} \in \mathbb{R}^3$, we denote an individual kernel center position and its corresponding support is abbreviated as $V_x := \textit{supp} \{ W_h \}$. Positions of specific Lagrangian particles are represented by $\mathbf{y} \in \mathbb{R}^3$. Further, we assume a constant kinematic viscosity $\nu = const$ and Mach number $Ma < 0.3$. The latter implies that we can simplify the viscous stress term to $div [ \boldsymbol{\tau}_{visc}  ]= div [ 2\nu \rho \boldsymbol{D} ] $ as $\nabla \cdot \mathbf{v} \simeq 0~$ \cite{Landau_1991, Jakobsen_2014}, with $\boldsymbol{D}$ denoting the symmetric strain rate tensor. Finally, the considered LES model in its Lagrangian form reads 
\begin{widetext}
\begin{subequations}
\label{eq:LES_Model}
\begin{eqnarray}
 \overline{\rho} (\mathbf{x}, t) & = & \int\displaylimits_{V_x} \rho(\mathbf{y},t)  W_h(\mathbf{x} - \mathbf{y})~\mathrm{d}\mathbf{y}~, \label{eq:LES_Mass} \\
    \overline{\rho} \frac{\mathrm{d} \tilde{\mathbf{v}}}{\mathrm{d}t}  (\mathbf{x}, t) & = &  -\int\displaylimits_{V_x} \nabla_\mathbf{y} p(\mathbf{y},t) W_h(\mathbf{x} -  \mathbf{y})~\mathrm{d}\mathbf{y} + 
    \int\displaylimits_{V_x} div_\mathbf{y} [ 2\nu \rho \boldsymbol{D}](\mathbf{y}, t) W_h(\mathbf{x} -  \mathbf{y})~\mathrm{d}\mathbf{y} 
    - div_{\mathbf{x}} \left[ \boldsymbol{\tau}_{SFS} \right](\mathbf{x}, t) ~, \label{eq:LES_Momentum} \\
    \overline{p} (\overline{\rho})  & = & p_{ref} + K \left( \frac{\overline{\rho}}{\rho_{ref}} - 1 \right), \quad \rho_{ref}, p_{ref}, K \in \mathbb{R}^+~.    \label{eq:LES_EOS}
\end{eqnarray}
\end{subequations}
\end{widetext}
The equations Eqs. (\ref{eq:LES_Mass}), (\ref{eq:LES_Momentum}) \& (\ref{eq:LES_EOS}) represent the averaged continuity equation \cite{Okraschevski_2021_2}, the averaged momentum transport equation and a linear barotropic equation of state (EOS) with $\rho_{ref},~p_{ref}~\&~K$ as constants. These describe the reference density of the strongly subsonic flow, the reference pressure and a stiffness constant, which are highly dependent on the problem. Their choice will be specified in Sec. \ref{sec:HIT}. Furthermore, from Eq. (\ref{eq:LES_Mass}), the meaning of the overline notation for an arbitrary field $f$ can be deduced. It describes a spatial average over a superfluid element $V_x$ of size $\sim h$, namely
\begin{equation}
    \overline{f} (\mathbf{x}, t) := \int\displaylimits_{V_x} f(\mathbf{y}, t) W_h(\mathbf{x} - \mathbf{y})~\mathrm{d}\mathbf{y}
    \label{eq:Filter}
\end{equation}
with $\mathrm{d}\mathbf{y}$ as volume differential of a Lagrangian fluid element. Moreover, the momentum transport equation Eq. (\ref{eq:LES_Momentum}) employs a density-weighted averaged velocity $\tilde{\mathbf{v}}$ over $V_x$ as indicated by the tilde notation. Generally, this density-weighted average for a field $f$ is termed Favre average \cite{Garnier_2009, Bilger_1975}, although it was already suggested by Reynolds \cite{Reynolds_1895} in 1895. It is defined as
\begin{equation}
    \tilde{f} (\mathbf{x}, t) := \frac{\overline{\rho f} (\mathbf{x}, t)}{\overline{\rho} (\mathbf{x}, t)}~.
    \label{eq:Favre}
\end{equation}
The use of Favre averages is not mandatory but handy, as it circumvents correlation terms related to the density field \cite{Bilger_1975}. It is important to distinguish between quantities according to Eq. (\ref{eq:Filter}) \& Eq. (\ref{eq:Favre}), which refer to super fluid elements in the LES framework, and fluid element quantities, which are simply noted without an overline or tilde. Additionally, as a consequence of the coarse-graining regularization of the balance equations by Eq. (\ref{eq:Filter}), an extra term $div_{\mathbf{x}} \left[ \boldsymbol{\tau}_{SFS} \right](\mathbf{x}, t)$ appears in Eq. (\ref{eq:LES_Momentum}) \cite{Eyink_2018, Okraschevski_2021_2}. It is the contribution from scales below $V_x$ to the momentum transport of super fluid elements. The SFS tensor $\boldsymbol{\tau}_{SFS}$ can be written as covariance tensor of the velocity field \cite{Okraschevski_2021_2}
\begin{widetext}
\begin{equation}
    \boldsymbol{\tau}_{SFS} (\mathbf{x}, t) := \int\displaylimits_{V_x} \rho(\mathbf{y}, t) (\mathbf{v}(\mathbf{y}, t) - \tilde{\mathbf{v}}(\mathbf{x}, t))(\mathbf{v}(\mathbf{y}, t) - \tilde{\mathbf{v}}(\mathbf{x}, t))^T W_h(\mathbf{x} - \mathbf{y}) ~ \mathrm{d}\mathbf{y}  ~,
    \label{eq:SFSSuperFluid}
\end{equation}
\end{widetext}
which is an interesting representation as the discretized version of Eq. (\ref{eq:SFSSuperFluid}) localizes flow subdomains, where SPH struggles with accurate approximations \cite{Okraschevski_2021_1}. Due to its relevance for this study, we will elaborate on this in more detail in Sec. \ref{sec:SFS}. 

Finally, it might be surprising that the averages on the right hand side of the transport equations in Eq. (\ref{eq:LES_Model}) are explicitly noted for each $V_x$ and not abbreviated by Eq. (3). By that we intend to emphasize that we follow the philosophy of explicit LES methods, e.g. Refs. \cite{Bose_2010, Radhakrishnan_2012}. Contrary to the usual procedure in explicit LES, where the nonlinear convective term is explicitly filtered, the filter is explicitly applied to the right hand side of the transport equations. This is due to the Lagrangian perspective we take, in which the convective term is not directly considered but rather a consequence of the individual forces on the right hand side of Eq. (\ref{eq:LES_Momentum}).

\section{\label{sec:SPH} SPH as a Lagrangian Quadrature of Explicit LES}

As explained in the last section, the governing equations of LES can generally be derived by coarse-graining of Lagrangian fluid elements using the Hardy theory from NEMD. This explains the conceptual similarity of LES and SPH, which becomes also evident from FIG. \ref{fig:01_HardyLES}, being a reminder of how SPH is often vividly introduced, e.g. in the work of Price \cite{Price_2012}. 

The objective of this section is to argue that SPH should be generally viewed as a Lagrangian quadrature technique for the governing equations of explicit LES. This general fluid dynamic framework includes the kernel concept from the beginning \textcolor{black}{and requires identical to SPH a joint limit for formal convergence \cite{Zhu_2015}, in which the ratio of filter width to grid spacing $\Delta/\Delta l \to \infty$ and  $\Delta \to 0$ \cite{Radhakrishnan_2012}}. In the following, we will present the resulting SPH model and discuss the implications of the explicit LES perspective.

\subsection{\label{subsec:DiscretizedLES} The SPH-LES Model and its Implications}

Decomposing the fluid domain into a finite number of Lagrangian SPH particles $i \in \{1,~...,~N\}$ that are connected to the kernel center positions, i.e.  $\forall i \in \{1,~...,~N\}: \mathbf{x}_i =  \mathbf{y}_i$, one can derive the final SPH model. The discretization procedure of Eq. (\ref{eq:LES_Model}) is detailed in the Appendix \ref{sec:Appendix} (Eqs. (\ref{eq:SPHDensity_2}), (\ref{eq:SPHPressureGrad_7}) \& (\ref{eq:SPHViscDiv_6})). It is important to highlight that the SPH particles only have to be Lagrangian representatives of super fluid elements $V_x$ with the arbitrary length scale $h$ instead of fluid elements. This is a significant difference to the usual SPH approach because traditionally SPH particles suffer from pseudo-Lagrangian behaviour at finite resolution \cite{Vogelsberger_2012, Okraschevski_2021_1}. For an individual particle $i$ with $j \in \{1,~...,~N_{ngb}\}$ neighbors the model reads
\begin{widetext}
\begin{subequations}
\label{eq:LES_Quadrature}
\begin{eqnarray}
 \overline{\rho}_i & = & M_i \sum_{j=1}^{N_{ngb}}  W_{h,ij} = \frac{M_i}{V_i} \quad \& \quad V_i := \frac{1}{\sum_{j=1}^{N_{ngb}}  W_{h,ij}}~, \label{eq:LES_Mass_Quadrature} \\
    \overline{\rho}_i \frac{\mathrm{d} \tilde{\mathbf{v}}_i}{\mathrm{d}t} & = &  - \sum_{j=1}^{N_{ngb}} ( \overline{p}_j + \overline{p}_i ) \nabla W_{h,ij} V_j + 
    2(2+n) \eta  \sum_{j=1}^{N_{ngb}}  \frac{(\tilde{\mathbf{v}}_i - \tilde{\mathbf{v}}_j ) \cdot (\mathbf{x}_i -  \mathbf{y}_j)}{(\mathbf{x}_i - \mathbf{y}_j)^2} \nabla W_{h,ij} V_j 
    - div\left[ \boldsymbol{\tau}_{SFS} \right]_i ~, \label{eq:LES_Momentum_Quadrature} \\
    \overline{p}_i (\overline{\rho}_i)  & = & p_{ref} + K \left( \frac{\overline{\rho}_i}{\rho_{ref}} - 1 \right), \quad \rho_{ref}, p_{ref}, K \in \mathbb{R}^+~,    \label{eq:LES_EOS_Quadrature}
\end{eqnarray}
\end{subequations}
\end{widetext}
and the particle trajectories follow from the kinematic condition
\begin{equation}
    \frac{\mathrm{d} \mathbf{x}_i}{\mathrm{d}t} = \tilde{\mathbf{v}}_i~.
    \label{eq:Kinematics}
\end{equation}
Formally, the emerging system of Eqs. (\ref{eq:LES_Quadrature}) \& (\ref{eq:Kinematics}) is identical to the SPH discretization of the weakly-compressible Navier-Stokes equations (WCSPH) except for the SFS term $div\left[ \boldsymbol{\tau}_{SFS} \right]_i$ in Eq. (\ref{eq:LES_Momentum_Quadrature}). The latter is a direct consequence of the coarse-graining at the arbitrary kernel scale $h$, compensating for subkernel effects. Contrary, in traditional SPH, the choice of the scale $h$ is merely a matter of convergence. We understand this as a physically convincing argument, going beyond empty formalities, to state that SPH should be understood as a Lagrangian quadrature technique intrinsically connected to explicit LES. Then, from this LES perspective, deficits introduced at the kernel scale for a specific choice of $h$ could potentially be compensated by a proper modelling of the SFS tensor $\boldsymbol{\tau}_{SFS}$ in Eq. (\ref{eq:LES_Momentum}). We believe that empirical evidence for this reconsideration is also given by the fact that already in the pioneering SPH works of Lucy and Gingold \& Monoghan \cite{Lucy_1977, Gingold_1977} artificial damping terms were used. These can be interpreted as the first SFS models accounting for subkernel deficiencies. Eventually, the reinterpretation of SPH as an intrinsic Lagrangian quadrature of explicit LES comes with two significant implications:

\begin{enumerate}
    \item \underline{Implication}: Ideally, the physical resolution of SPH is limited by the kernel scale of $V_x$ or more precisely the kernel diameter $D_K$. Thus, SPH is unsuited as a DNS method.
    \item \underline{Implication}: Deficits introduced below the kernel scale of $V_x$ might be resolved by explicit consideration of the SFS term, from which structures above the kernel scale of $V_x$ could profit.
\end{enumerate}

While the first implication can be easily understood and there is already empirical evidence proving it, e.g. the work of Bauer \& Springel \cite{Bauer_2012}, the second implication should be interpreted as a working hypothesis, which we will test by numerical experiments in Sec. \ref{sec:Results}. However, before this, \textcolor{black}{we will recapitulate the current knowledge about the origin of numerical dissipation in SPH in the next section and explain its relation to the SFS.}

\subsection{\label{sec:SFS} \textcolor{black}{Numerical Dissipation and the Role of the SFS}}

\begin{figure*}[t]
\includegraphics[width=6.5in]{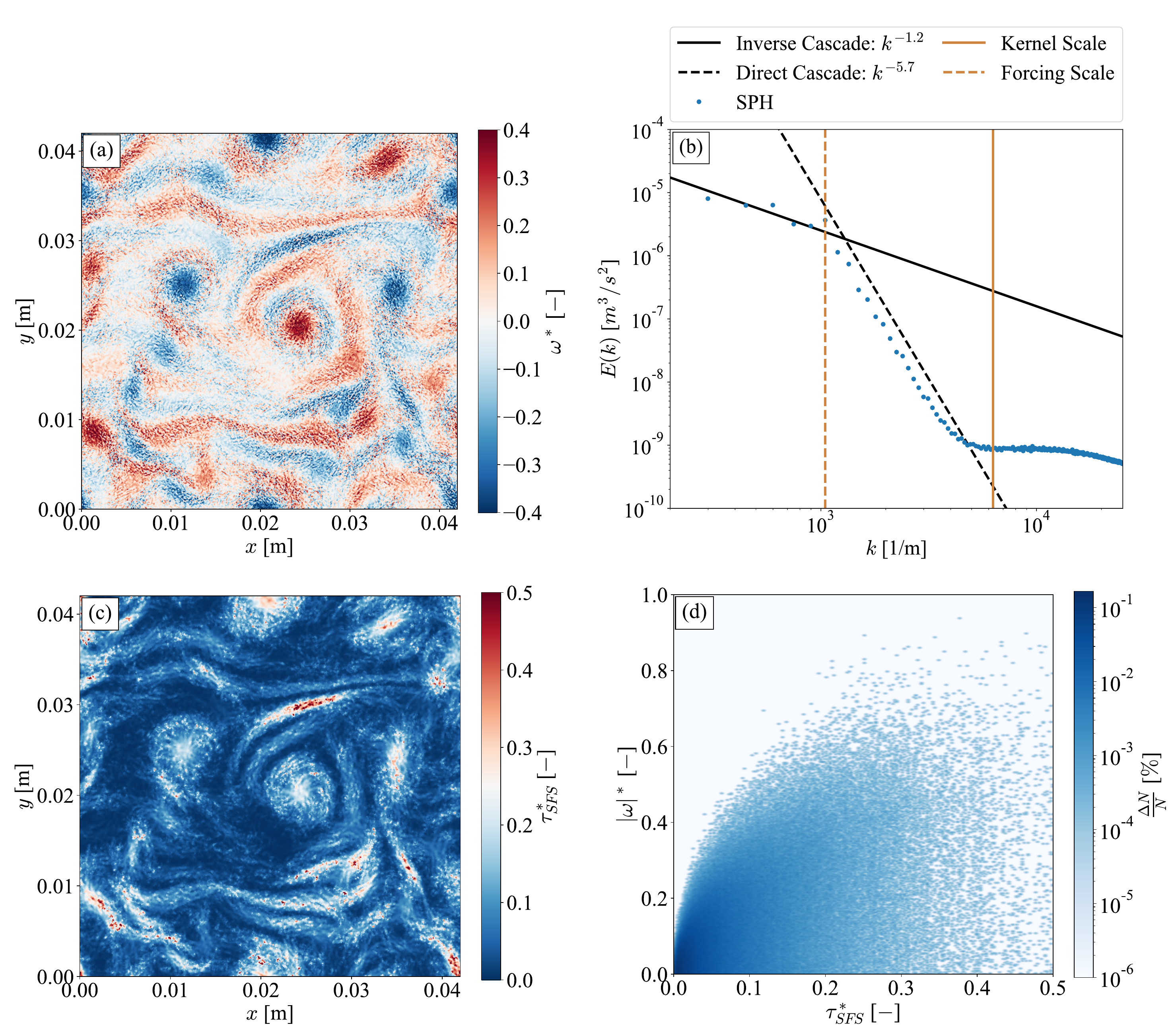}
\caption{Metrics of a SPH solution for a two dimensional turbulent Kolmogorov flow according to Rivera \emph{et al.} \cite{Rivera_2000, Rivera_2016}. (a) Snapshot of the nondimensional, noisy vorticity field, (b) Kinetic energy spectrum, (c) Snapshot of the nondimensional Frobenius norm of the SFS tensor and (d) Nondimensional bivariate probability density of the Frobenius norm of the SFS tensor (abscissa) and vorticity (ordinate).}
\label{fig:02_KolmogorovSPHStress}
\end{figure*}

\textcolor{black}{In this paragraph, we will summarize how numerical dissipation in kernel-based particle methods like SPH emerges and how it can be localized by usage of Hardy theory again. Exemplary, we explain the coherencies with the aid of a SPH solution of the forced two dimensional turbulent Kolmogorov flow of Rivera \emph{et al.} \cite{Rivera_2000, Rivera_2016}. This was the main topic of our work in Ref. \cite{Okraschevski_2021_1}, which subsequently is summarized. As the discussion will reveal, the current understanding of numerical dissipation in SPH is rather heuristic than rigorous compared to conventional grid-based CFD methods. The most important metrics for the analyzed Kolmogorov flow are depicted in FIG. \ref{fig:02_KolmogorovSPHStress} and non-dimensionalized with the absolute maximum value of the viewed snapshot. }

\textcolor{black}{Although the Kolmogorov flow can be generally reproduced by proper calibration with particle discretization methods in terms of the energy characteristics, the analysis shows that the dissipation rate is strongly overpredicted. From current understanding, the excessive numerical dissipation is rooted in increasing particle disorder at higher Reynolds number causing local SPH particle transport perpendicular to the main flow direction \cite{Ellero_2010}. The resulting unphysical momentum transport prevails especially in zones with large velocity gradients and introduces artificial vorticity fluctuations \cite{Colagrossi_2013}. For the considered problem, exemplary evidence is given by the noisy vorticity field in FIG. \ref{fig:02_KolmogorovSPHStress} (a). Since the turbulent flow is periodic, it can be analytically argued that the vorticity fluctuations $\omega '$ will quadratically increase the averaged dissipation rate $\epsilon$ according to \cite{Bailly_2015} }
\begin{equation}
    \epsilon = \nu \langle \omega '^2 \rangle_V = 2 \nu \int_0^\infty k^2 E(k)~\mathrm{d}k~,
    \label{eq:Dissipation}
\end{equation}
\textcolor{black}{with $\langle \cdot \rangle_V$ denoting a volume average. Moreover, based on Eq. (\ref{eq:Dissipation}), the vorticity fluctuations, causing numerical dissipation, are evidently linked to the kinetic energy spectrum $E(k)$. An analysis of the extracted spectrum in FIG. \ref{fig:02_KolmogorovSPHStress} (b) for the Kolmogorov flow demonstrates that the scalings of the inverse and direct cascades known from the experiment can be matched above the kernel scale ($k < 2\pi/D_K$) \cite{Rivera_2016}. Merely the saturation of $E(k)$ below the kernel scale ($k > 2\pi/D_K$) deviates from the ideal form. Since dissipation takes place at small scales due to the $\sim k^2$ weighting of $E(k)$ in Eq. (\ref{eq:Dissipation}), it seems likely that this saturation represents the spectral signature of numerical dissipation. In accordance with FIG. \ref{fig:00_SchematicSpectra}, we will term this bottleneck as artificial thermal range or artificial thermalization. It should be emphasized that this spectral signature was also observed in other works, e.g. Refs. \cite{Bauer_2012, DiMascio_2017, Antuono_2021}.}

\textcolor{black}{The upper paragraph should be understood as a heuristic description of numerical dissipation in kernel-based particle methods as SPH. However, it does not locally explain how numerical dissipation in the flow field takes place. Motivated by this deficit, we applied Hardy theory to a numerical particle set in Ref. \cite{Okraschevski_2021_1} deriving a tensor, namely a discrete approximation of the SFS tensor, which was shown to serve this purpose. It reacts to large velocity gradients and particle noise, which are believed to be the root of the numerical dissipation. In Ref. \cite{Okraschevski_2021_1} the tensor was termed as molecular stress, since we were not aware of its connection to explicit LES. Interestingly, in kernel-based particle methods this tensor can always be estimated, even if no LES perspective is employed. Considering the fact that two different spatial resolution scales exist, i.e. the kernel diameter $D_K$ and the particle size $\Delta l < D_K$, a Lagrangian quadrature of Eq. (\ref{eq:SFSSuperFluid}) gives the following estimate}
\begin{equation}
    \boldsymbol{\tau}_{SFS,i}  \approx \sum_{j=1}^{N_{ngb}} \overline{\rho}_j ( \tilde{\mathbf{v}}_j - \tilde{\mathbf{v}}_i ) ( \tilde{\mathbf{v}}_j - \tilde{\mathbf{v}}_i )^T W_{h,ij} V_j ~. 
    \label{eq:SPH_SFS_6}
\end{equation}
\textcolor{black}{For the considered Kolmogorov flow, an evaluation of the Frobenius norm of the SFS tensor according to Eq. (\ref{eq:SPH_SFS_6}) clearly demonstrates that the SFS estimate fits into the former dissipation characteristics. A visual comparison between the vorticity (FIG. \ref{fig:02_KolmogorovSPHStress} (a)) and the SFS tensor (FIG. \ref{fig:02_KolmogorovSPHStress} (c)) reveals an unambiguous relationship, which is supported by the corresponding bivariate probability density function in FIG. \ref{fig:02_KolmogorovSPHStress} (d). The latter describes a cone like structure indicating that high levels of absolute vorticity and SFS are connected \cite{Okraschevski_2021_1}, as well as that the variance of the absolute vorticity increases with the SFS norm up to $\tau_{SFS}^* \lesssim 0.4$. Consequently, we conclude that the quantities $\omega '$, $E(k)$ \& $\boldsymbol{\tau}_{SFS}$ are evidently correlated with each other, containing different levels of information about the numerical dissipation dynamics.}

\textcolor{black}{With the former relations,} the role of the SFS term $div\left[ \boldsymbol{\tau}_{SFS} \right]_i$ in the discretized explicit LES equations (Eq. (\ref{eq:LES_Quadrature})) becomes apparent. Since the SFS term behaves diffusive regarding the velocity field in a statistically averaged sense \cite{Eyink_2018, Moser_2021}, an explicit consideration of this term will \textcolor{black}{attempt} to locally homogenize the velocity field. \textcolor{black}{Hence, a mitigation of the SFS norm according to Eq. (\ref{eq:SPH_SFS_6}) will be caused, which is expected to reduce the vorticity variance according to FIG. \ref{fig:02_KolmogorovSPHStress} (d). Then, based on Eq. (\ref{eq:Dissipation}), the artificial thermalization of the kinetic energy spectrum should be likewise reduced. Considering again that dissipation takes place at small scales due to the $\sim k^2$ weighting of $E(k)$ in Eq. (\ref{eq:Dissipation}), this could potentially enable a reduction of numerical dissipation from which large scale structures might profit. Thus, the main idea of this work is to replace numerical dissipation by an explicit, dissipative SFS model, which counteracts the numerically induced small scale fluctuations.}

To verify these \textcolor{black}{anticipated causalities}, we deem the eddy viscosity concept in connection with Boussinesq’s hypothesis \cite{Sagaut_2006, Schmitt_2007, Silvis_2017, Moser_2021} for the modelling of the SFS term $div\left[ \boldsymbol{\tau}_{SFS} \right]_i$ in Eq. (\ref{eq:LES_Momentum_Quadrature}) as adequate. The following approaches will be utilized:
\begin{itemize}
    \item \texttt{SMAG}: This represents the classical Smagorinsky model discretized according to Eqs. (\ref{eq:SPH_SFS_4}), (\ref{eq:SPH_SFS_7}) and (\ref{eq:SPH_SFS_10}). It is angular momentum conserving in the continuum limit \cite{Sijacki_2006}.
    \item \texttt{SIGMA}: This represents the superior $\sigma$-model of Nicoud \emph{et al.} \cite{Nicoud_2011} discretized according to Eqs. (\ref{eq:SPH_SFS_4}), (\ref{eq:SPH_SFS_7}) and (\ref{eq:SPH_SFS_11}). It is also angular momentum conserving in the continuum limit \cite{Sijacki_2006} but should overcome severe drawbacks of the Smagorinsky model, e.g. non-vanishing subfilter dissipation in laminar regions \cite{Silvis_2017, Nicoud_2011}. 
    \item \texttt{SMAG-MCG}: This represents the classical Smagorinsky model, however, discretized in the Monoghan-Cleary-Gingold (MCG) form \cite{Colagrossi_2017,Cleary_1999}. It is angular momentum conserving on the particle level as well.
\end{itemize}

It is of paramount importance that the SFS dissipation is introduced only on subkernel scales in order to guarantee a successful application of the explicit SFS model, eventually reducing the artificial thermalization. However, the SPH discretization requires non-local approximations, which might jeopardize this goal a priori. This can be vividly illustrated by the concept of \emph{Particle Duality}, which results from the coarse-graining perspective.

\subsection{\label{subsec:ParticleDuality} Particle Duality and Numerical Dispersion}

\begin{figure}
\includegraphics[width=3.4in, clip, trim= 0cm 0cm 0cm 0cm]{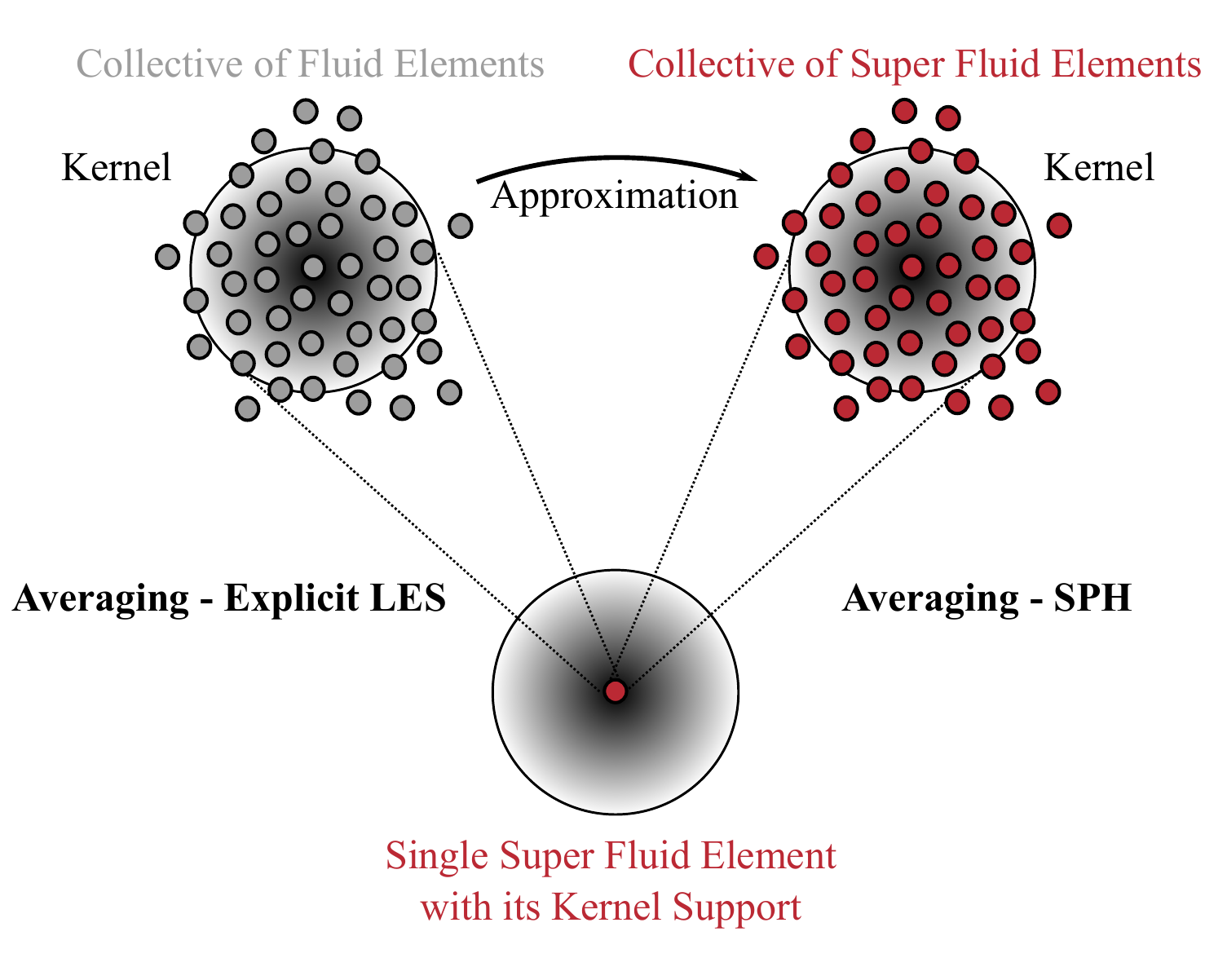}
\caption{Schematic to illustrate the concept of \emph{Particle Duality}. SPH particles represent super fluid element approximants and fluid element surrogates at the same time.}
\label{fig:03_ParticleDuality}
\end{figure}

In order to understand the concept of \emph{Particle Duality}, it is necessary to precisely define the terminology of explicit LES. According to Sec. \ref{sec:Turbulence} of this work, explicit LES is introduced as a general fluid dynamic framework, in which fluid elements are coarse-grained by an explicit kernel to so called super fluid elements. This is illustrated in the left part of the schematic in FIG. \ref{fig:03_ParticleDuality}. From the schematic, an averaging over a fluid element collective (grey particles) exactly determines the properties of a single super fluid element (red particle) with its specific kernel support. This corresponds to a truly explicit LES. However, in a SPH model the fluid element properties are unknown, which is synonymous to the closure problem of turbulence. This issue is resolved in a SPH framework by a direct substitution of the fluid elements (grey particles) by super fluid elements (red particles). Only then an averaging is performed to estimate the properties of a single super fluid element itself. As a consequence the SPH particles must represent super fluid element approximants and fluid element surrogates at the same time, which is what we term as \emph{Particle Duality}. Practically, this implies that super fluid element approximants interact with each other, which are not direct neighbors but rather separated by some particles in between. This occurs as long as a the particles share the same kernel support and implicitly causes an increase of the effective interaction distance. Physically, however, the considered interaction is inadequate as the governing LES equations are a local model in terms of the super fluid element quantities. Thus, the \emph{Particle Duality} as a manifestation of the non-locality introduced by the SPH discretization gives an picturesque description why the consideration of an explicit dissipative SFS model might fail to result in an improvement as anticipated in Sec. \ref{sec:SFS}. Conceivably, the SFS model will not solely remove kinetic energy from the problematic artificial thermal range but also affect resolved scales larger than the kernel. 

\textcolor{black}{From the phenomenological issue described above, one might not exclusively expect problems in terms of the application of dissipative SFS models. It is also likely that numerical dispersion effects become relevant. Since the \emph{Particle Duality} results in a decreased effective resolution, it should be anticipated that modes of a certain wavenumber $k$ will be properly resolved only if the kernel wavenumber $k_{kern}$ exceeds $k$ significantly, i.e. $k \ll k_{kern}$, and $N_{ngb} \to \infty$. Indeed generic SPH dispersion studies for the Euler equations demonstrate that the resulting dispersion errors depend on $(k/k_{kern})$, the kernel type, the number of neighbors $N_{ngb}$ inside the kernel, the reference density $\rho_{ref}$ and pressure level $p_{ref}$, e.g. Ref. \cite{Dehnen_2012}. However, from these insights it can be hardly predicted how turbulent flows will be affected. A recent study of Yalla et al. \cite{Yalla_2021} investigates the effect of numerical dispersion effects on the energy cascade in grid-based LES. The authors conclude that numerical dispersion can spoil the Galilean invariance in numerical simulations, finally inhibiting the energy transfer to the small, most dispersive modes, hence causing a degradation of the inertial range in the energy spectrum. Though the one-to-one transfer of these results to Lagrangian kernel-based particle methods as SPH is questionable, we will still test in Sec. \ref{sec:Results} whether Galilean invariance holds. This might give an indication about the relevance of numerical dispersion errors in our study. }

Having laid the foundations of SPH as a discretization intrinsically connected to explicit LES, it is now indispensable to answer our central question: \emph{Can resolved large scale structures profit from the reduction of SPH typical small scale noise by explicit use of SFS models}? Therefore, a thorough investigation of the influence of explicit SFS models for a well-defined HIT problem are vital, which is why we proceed with the description of such in the next section.

\section{\label{sec:HIT} The HIT Problem}

\begin{figure*}[t]
\includegraphics[width=6.5in, clip, trim= 0cm 3cm 0cm 3cm]{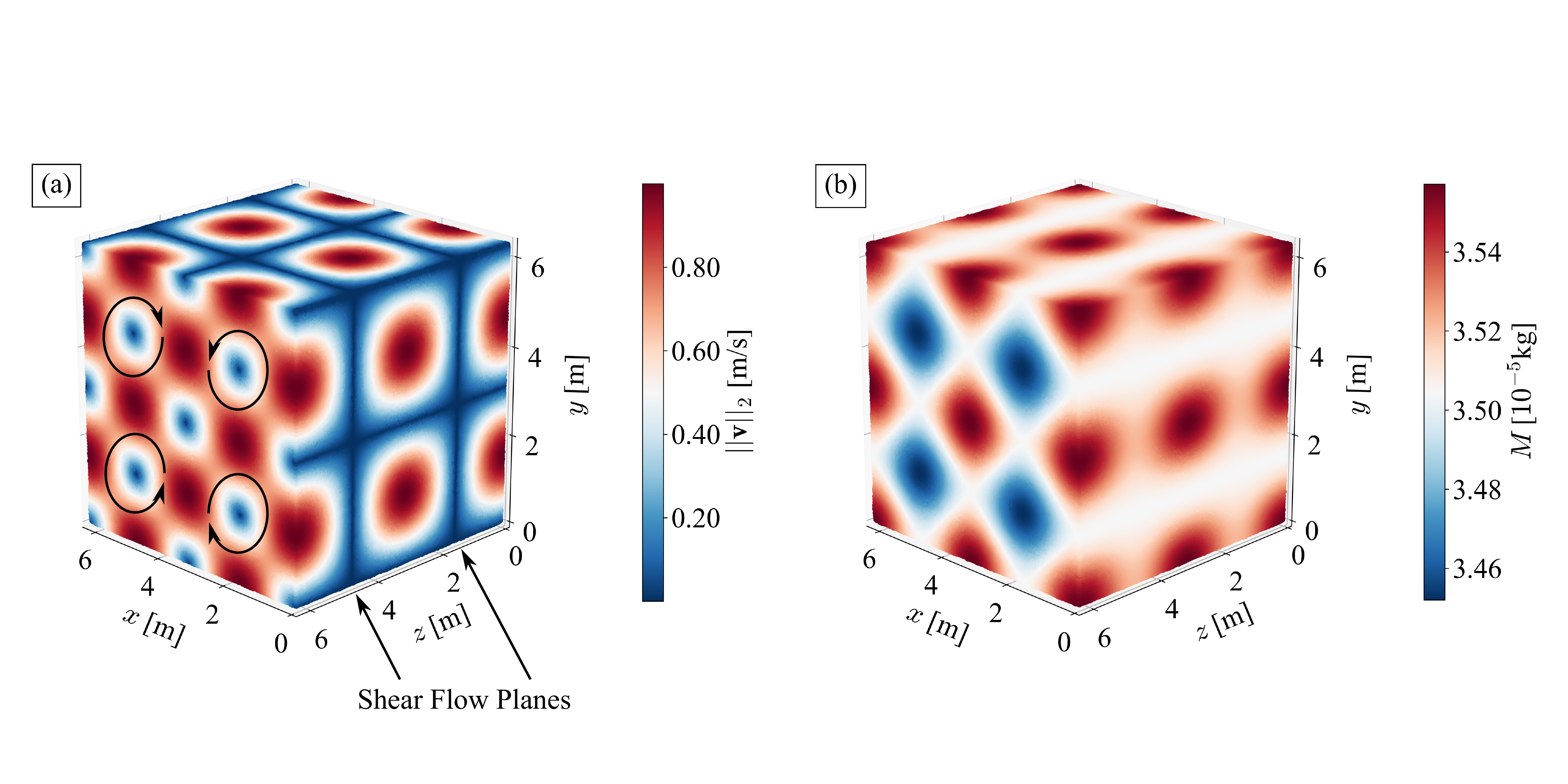}
\caption{Initialization of the Taylor-Green flow at $Re=10^4$ with $N=192^3$ particles. (a) Velocity magnitude field with highlighted shear flow planes and rotational direction for $z=2\pi$. (b) Mass distribution corresponding to the initial pressure field.}
\label{fig:04_InitialHIT}
\end{figure*}

The HIT problem that will be subsequently investigated is the Taylor-Green flow \cite{Taylor_1937, Brachet_1983} at $Re=10^4$ presented in the work of Dairay \emph{et al.} \cite{Dairay_2017} (\textcolor{black}{other values of $Re$ are discussed in Appendix \ref{sec:AppendixB}}). \textcolor{black}{Their DNS solution will serve as reference for our study, as well as an second order accurate (space \& time) solution of a standard Smagorinsky LES from a Finite Volume Methode (FVM). The latter was computed with OpenFOAM 7 on a Cartesian grid with $384^3$ cells and closely matches the energy characteristics of the DNS.} The initial velocity field in the tri-periodic domain $\Omega := [0,2\pi]^3$ of the freely decaying flow is specified as
\begin{eqnarray}
    &v_{0,x} (x,y,z) =& ~\mathrm{sin}(x)\mathrm{cos}(y)\mathrm{cos}(z) \nonumber \\
    &v_{0,y} (x,y,z) =& -\mathrm{cos}(x)\mathrm{sin}(y)\mathrm{cos}(z) \label{eq:TGVel} \\
    &v_{0,z} (x,y,z) =& ~0 \nonumber
\end{eqnarray}
and the corresponding pressure field follows from the solution of the pressure Poisson equation in the incompressible limit \cite{Pereira_2021}
\begin{equation}
    p_{0} (x,y,z) = p_{ref} + \frac{\rho_{ref}v_0^2}{16} ( 2 + \mathrm{cos}(2z) ) ( \mathrm{cos}(2x) + \mathrm{cos}(2y) )~
    \label{eq:TGPress}
\end{equation}
\textcolor{black}{with $v_0=1~\mathrm{m/s}$.} The SPH cases that will be presented in the following are summarized in TABLE \ref{tab:Cases}. Generally, four different particle counts were considered, namely $N \in \{128^3,~192^3,~256^3,~512^3\}$, ranging from $\sim 2$ Mio. particles to $\sim 130$ Mio. particles. The basis of the given particle powers define the averaged particle distance $\Delta l$, which exemplary for Case 1 results in $\Delta l = 2\pi/128~ \mathrm{m} \approx 0.0491~\mathrm{m}$. Starting from a Cartesian lattice arrangement, the particles were regularized into a stable configuration in corresponding pre-runs following the particle packing scheme of Colagrossi \emph{et al.} \cite{Colagrossi_2012}. Only then, the fields given by Eqs. (\ref{eq:TGVel}) \& (\ref{eq:TGPress}) were mapped onto the particles. The initial velocity field magnitude for a $N=192^3$ Case is depicted in FIG. \ref{fig:04_InitialHIT} (a) highlighting two shear flow planes at $z=\pi/2$ \& $z=3\pi/2$ and the rotational direction at the plane $z=2\pi$. In order to match the pressure field in the initial time step and avoid artificial dynamical effects beyond the one resulting from the initial particle configuration, a consistent mass distribution $M_{0,i} = \rho_{0,i}\Delta l^3$ was imposed. It is illustrated in FIG. \ref{fig:04_InitialHIT} (b). The density field $\rho_{0,i}$ is given by the combination of the initial pressure field in Eq. (\ref{eq:TGPress}) and the EOS in Eq. (\ref{eq:LES_EOS_Quadrature}). For the latter a reference density of $\rho_{ref} = 1 ~\mathrm{kg/m^3}$, a stiffness constant $K=\rho_{ref}c_a^2=25~\mathrm{Pa}$ and a reference pressure $p_{ref} = K/4 = 6.25~\mathrm{Pa}$ were chosen. The stiffness constant implies an artificial speed of sound of $c_a=5~\mathrm{m/s}$, which corresponds to an initial Mach number $Ma_0 = 0.2$ and justifies to neglect $\nabla \cdot \mathbf{v}$ based forces \cite{Jakobsen_2014}. We want to emphasize that different values of $p_{ref}$ were tested, namely $p_{ref} \in \{ K/10,~K/4,~K/2,~K \}$, however, the value $p_{ref} = K/4$ yielded the best trade-off between stability and numerical dissipation. The results in the following were all computed using a Wendland C4 kernel and a kernel diameter of $D_K=8\Delta l$ resulting in $N_{ngb} \approx 250$. Other comparative simulations were conducted with $D_K \in \{ 4,~6,~8 \}~\Delta l$ (\textcolor{black}{see Appendix \ref{sec:AppendixC}}) and a quintic B-spline kernel but only the chosen configuration was capable to provide reasonable numerical convergence with increasing $N$ avoiding pairing instabilities at the same time \cite{Zhu_2015, Dehnen_2012}. \textcolor{black}{If not mentioned otherwise, the before described strategy will be used for all runs in this study.}

\begin{table}[t]
\renewcommand{\arraystretch}{1.2}
\caption{\textcolor{black}{SPH cases for the considered Taylor-Green flow.}}
\label{tab:Cases}
\begin{ruledtabular}
\begin{tabular}{cccccccc}
    Case & \multicolumn{4}{c}{Particles} & \multicolumn{3}{c}{SFS Model} \\
     & $128^3$ & $192^3$ & $256^3$ & $512^3$ & \texttt{SMAG} & \texttt{SIGMA} & \texttt{SMAG-MCG}\\

\hline
    1 & $\times$ & & & & & &   \\
    2 & & $\times$ & & & & &   \\
    3 & & & $\times$ & & & &   \\
    4 & & & & $\times$ & & &   \\
\hline
    5 & $\times$ & & & & $\times$ & &   \\
    6 & & $\times$ & & & $\times$ & &   \\
    7 & & & $\times$ & & $\times$ & &   \\
    8 & & & & $\times$ & $\times$ & &   \\
\hline
    9 & & & $\times$ & & & $\times$ &   \\
    10 & & & $\times$ & & & & $\times$   \\
    11 & & & $\times$ & & $C_S = 0.075$ & &  \\
    12 & & & $\times$ & & $C_S=0.3$ & &  \\
\hline
    13 & & & $v_{0,z}=\pi$ & & & &
    
\end{tabular}
\end{ruledtabular}
\end{table}

To facilitate the discussion of explicit SFS models for the SPH method, the results in Sec. \ref{sec:Results} will follow the subsequent argumentation sequence: First, we will demonstrate that the system of Eqs. (\ref{eq:LES_Quadrature}) \& (\ref{eq:Kinematics}), shows a convergent tendency in the numerical sense (TABLE \ref{tab:Cases}: Case 1-4). These runs correspond to usual WCSPH simulations or, from our coarse-graining perspective in Sec. \ref{sec:SPH}, to a Lagrangian quadrature of the explicit LES equations (Eq. (\ref{eq:LES_Model})) without an explicit SFS model. Hence, the SFS will only be implicitly considered. The Cases 5-12 will represent runs in which the SFS is explicitly added by means of eddy viscosity approaches (Sec. \ref{sec:SFS}). To proof that our conclusions are generally valid for for eddy viscosity approaches, not only the standard Smagorinsky \texttt{SMAG} (TABLE \ref{tab:Cases}: Case 5-8) model will be evaluated but also the $\sigma$-model \texttt{SIGMA} (TABLE \ref{tab:Cases}: Case 9), which should ensure vanishing subfilter dissipation in the initial laminar phase of the HIT problem. Moreover, the  Smagorinsky model will be additionally combined with the angular momentum conserving MCG form in Eq. (\ref{eq:SPH_SFS_12}) called \texttt{SMAG-MCG} (TABLE \ref{tab:Cases}: Case 10), demonstrating the robustness of our observations. \textcolor{black}{In order to refute that the obtained results are the consequence of a wrong calibration of the model constant in the eddy viscosity models \cite{Moser_2021}, e.g. $C_S$ for Smagorinsky, we will further present two \texttt{SMAG} runs in which the standard value $C_S=0.15$ was either halved or doubled (TABLE \ref{tab:Cases}: Case 11-12).} All eddy viscosity runs were performed with a filter width $\Delta=D_K/2=R_K$ being equivalent to the kernel radius $R_K$ and in accordance with Rennehen \cite{Rennehen_2021}. From our explicit LES perspective in Sec. \ref{sec:SPH} the most consistent choice would correspond to $\Delta = D_K$ but some tests led to the conclusion that only the overall dissipation is enhanced without any further physical improvements. Interestingly, for the problem considered, the choice $\Delta = \Delta l$ had a nearly negligible effect on our solutions. We interpret this as evidence in favor of the intrinsic connection between explicit LES and SPH, in which the particles should approximate LES super fluid elements and not fluid elements itself. \textcolor{black}{This will be confirmed in Sec. \ref{subsec:OtherSFSModels}.} 

\textcolor{black}{The final Case 13 is identical to Case 3 except for the fact that in Eq. (\ref{eq:TGVel}) a constant velocity is superimposed on the third component, namely $v_{0,z} (x,y,z) = ~\pi$. By that we intend to test, according to Sec. \ref{subsec:ParticleDuality}, whether numerical dispersion might be of relevance for our investigation. If so, the Galilean invariance of the model could be spoiled, finally altering the shape of the spectrum obtained in Case 3 according to Yalla et al. \cite{Yalla_2021}. We want to note that the choice $v_{0,z} (x,y,z) = ~\pi$ is an order of magnitude larger than the velocity fluctuations at time $t=14~\mathrm{s}$, on which our analysis mostly focuses on. This implies a turbulence intensity of $\mathcal{O}(Tu)\approx 0.1$. Moreover, the maximum of the velocity magnitude in Case 13 corresponds to $v_{max}=\sqrt{1+\pi^2} \approx 3.3~\mathrm{m/s}$. To ensure initial Mach similarity with Case 1, a speed of sound $c_a=16.5~\mathrm{m/s}$ is required. This also alters the initial mass distribution at the beginning according to the strategy above and the choice of the reference pressure $p_{ref} = \rho_{ref} c_a^2/4 \approx 68~\mathrm{Pa}$ to ensure stability.}

To evaluate the quality of the results, different metrics will be invoked. On the one hand the assessment of the overall dissipation inside the domain will be based on the density weighted averaged kinetic energy 
\begin{equation}
    e_v(t) := \frac{1}{2} \frac{\sum_{i=1}^N \overline{\rho}_i \tilde{\mathbf{v}}_i^2 V_i}{\sum_{i=1}^N \overline{\rho}_i V_i}~.
    \label{eq:MassAverKineticEnergy}
\end{equation}
This metric is in accordance with the definition of Dairay \emph{et al.} \cite{Dairay_2017} except for the density weighting. It only has a minor influence as we could verify, but should be included for consistency with the weak compressibilty approach. On the other hand the overall dissipation will be assessed by the corresponding averaged dissipation rate, which can be computed from a finite difference approach for sufficient temporal sampling \cite{Dairay_2017} using the relation
\begin{equation}
    \epsilon_t (t) = - \frac{\mathrm{d} e_v}{\mathrm{d}t}~.
    \label{eq:DissipationRate}
\end{equation}
Furthermore and most importantly, we will compute the kinetic energy spectra $E(k)$ at time $t=14~\mathrm{s}$ where HIT with the characteristic inertial range scaling of Eq. (\ref{eq:HIT_Energy_Scaling}) should be present up to a wavenumber of $k_{DNS} \approx 50~\mathrm{1/m}$, \textcolor{black}{according to} Ref. \cite{Dairay_2017}. Therefore, we employ the nearest neighbor sampling technique of Bauer \& Springel \cite{Bauer_2012} on a Cartesian grid with $\Delta l/2$ in combination with the method of Durran \emph{et al.} \cite{Durran_2017}. This methodology is kinetic energy conserving or in other words satisfies the discrete Parseval relation. Due to the Nyquist criterion, spectra will only be presented up to wavenumbers corresponding to $2\Delta l$. The interpolation method of Shi \emph{et al.} \cite{Shi_2013} will not be considered as is unclear whether it might introduce smoothing in the artificial thermal range, which we want to avoid. Since the values $e_v(t=14~\mathrm{s})$ for the different cases in TABLE \ref{tab:Cases} can significantly differ, we will normalize the corresponding spectra with the product $e_v(t=14~\mathrm{s})L_c$ and $L_c=1~\mathrm{m}$ for all runs to enable a relative comparison. \textcolor{black}{Observations in spectral space will further be related to physical space by means of the Frobenius norm of Eq. (\ref{eq:SPH_SFS_6}), namely $||\boldsymbol{\tau}_{SFS}||_F$, the backward finite-time Lyapunov exponent (FTLE) \cite{Sun_2016, Dauch_2018} in the time range $[11,14]~\mathrm{s}$ and the vorticity component $\omega_z=\partial_x v_y - \partial_y v_x$}. While the first is indicative for small scale structures (see Sec. \ref{sec:SFS}), the other will be used to assess the large scale structures. Additionally, we introduce a signal to noise ($SNR$) metric for the kinetic energy spectra defined by the ratio of energy above the kernel scale (with kernel wavenumber $k_{kern}$) in relation to the overall energy, namely
\begin{equation}
    SNR := \frac{\int_{k=0}^{k_{kern}} E(k) \mathrm{d}k}{\int_{k=0}^{\infty} E(k) \mathrm{d}k}~.
    \label{eq:SNR}
\end{equation}
It is important to stress that the $SNR$ metric is only indicative for the reduction of the artificial thermalization but gives no insight about the solution quality above the the kernel scale.

All computations in this work were performed for a time range of $I_T=[0;15]~\mathrm{s}$ with the in-house SPH code \emph{turboSPH}. The latter was developed for the prediction of primary atomization. For details, please refer to the work of Chaussonnet \emph{et al.} \cite{Chaussonnet_2020}.

\section{\label{sec:Results} Results \& Discussion}

\subsection{\label{subsec:WCSPH} Numerical Convergence of the WCSPH Scheme}

\begin{figure*}
\includegraphics[width=5.7in, clip, trim= 0cm 0cm 0cm 0cm]{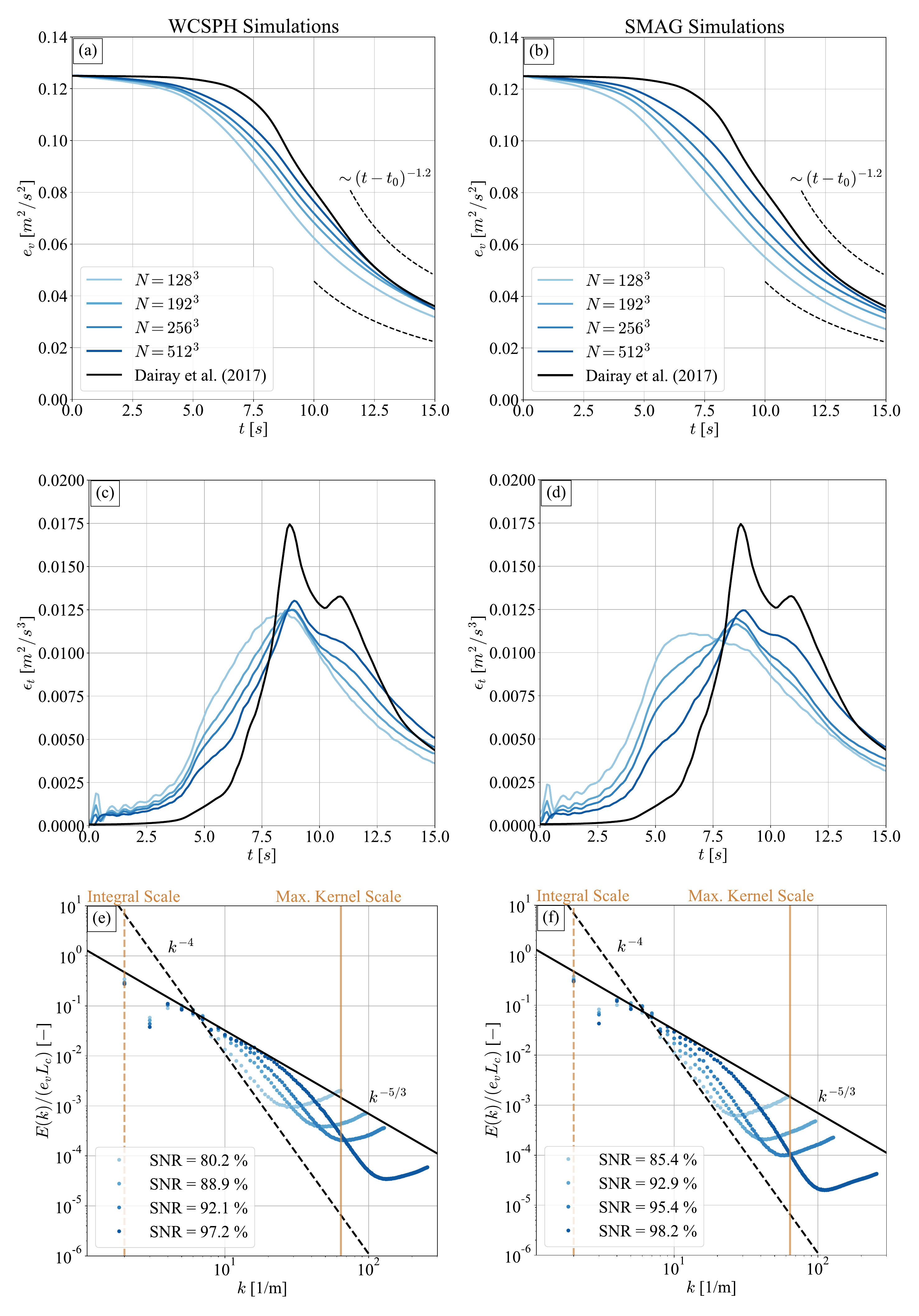}
\caption{\textcolor{black}{Comparison of quantitative metrics for different particle counts $N$. The first column (a), (c) \& (e) represents the WCSPH solutions without SFS model and the second column (b), (d) \& (f) represents the solution with \texttt{SMAG} SFS model. The color coding is explained in (a). (a) \& (b) Temporal evolution of the density
weighted averaged kinetic energy $e_v$. (c) \& (d) Temporal evolution of the averaged dissipation rate $\epsilon_t$. (e) \& (f) Kinetic energy spectra at $t=14~\mathrm{s}$.}}
\label{fig:05_EnergyBalance}
\end{figure*}

We will first start with a qualitative discussion of the numerical convergence of the discretized LES equations resulting from the Lagrangian quadrature (Eq. (\ref{eq:LES_Quadrature})). For now the SFS term is neglected, formally leading to the standard WCSPH discretization of the Navier-Stokes equations  (TABLE \ref{tab:Cases}: Case 1-4). Hence, these results will serve as reference to evaluate the effect of explicit SFS models in SPH. 

From FIG. \ref{fig:05_EnergyBalance} (a), which illustrates the temporal evolution of the averaged kinetic energy $e_v$ (Eq. (\ref{eq:MassAverKineticEnergy})) for increasing particle counts $N$ (darker colors), one can apparently conclude that the metric is numerically converging towards the DNS reference solution of Dairay \emph{et al.} \cite{Dairay_2017} (solid black line). However, even for $N=512^3 \approx 130~\mathrm{Mio.}$ particles, a significant gap remains compared to the DNS in the interval $t \in [2.5;~10]~\mathrm{s}$. The reason for this gap can be understood from the averaged dissipation rate $\epsilon_t$ (Eq. (\ref{eq:DissipationRate})) as depicted in FIG. \ref{fig:05_EnergyBalance} (c). Especially in the initial timeframe $t \in [0;~2.5]~\mathrm{s}$ the vanishing dissipation rate of the DNS solution is strongly overestimated by the WCSPH scheme. This is probably linked to the strong anisotropic particle rearrangement of the initially laminar vortex configuration, causing numerical dissipation effects. Interestingly, although the highest resolution contains $64$ times more particles than the lowest resolution, the curves only slowly approach the vanishing DNS level.
The overestimation of the dissipation rate continues until $t \approx 7.5~\mathrm{s}$, where the blue WCSPH lines cross the black DNS line, and passes over to a systematic underestimation until the end of the simulation, except for the $N=512^3$ case, which overestimates the dissipation rates again for $t \gtrsim 12.5~\mathrm{s}$. Nevertheless, this global dissipation characteristic leads to the consequence that the energy levels in FIG. \ref{fig:05_EnergyBalance} (a) approach the DNS solution again for $t \gtrsim 7.5~\mathrm{s}$. Despite these quantitative deviations, it should be highlighted that the qualitative agreement of the dissipation rates $\epsilon_t$ in FIG. \ref{fig:05_EnergyBalance} (c) is reasonable. For all $N$ the temporal occurrence of the dissipation peak at $t \approx 9~\mathrm{s}$ is matched and for increasing $N$ the formation of the second local dissipation peak at $t \approx 11~\mathrm{s}$ is also evident. \textcolor{black}{Moreover, it must be positively emphasized that after the first dissipation peak the dissipation rates are consistent with theoretical predictions \cite{Skrbek_2000}, which state that $e_v(t) \sim t^{-1.2}$. This is illustrated in FIG. \ref{fig:05_EnergyBalance} (a) by the two black dashed lines, which only differ in the time shift parameter $t_0$. The latter can be understood as an indicator for the begin of the decay process and is consequently smaller for the WCSPH runs than for the DNS.}

\begin{figure}[h]
\includegraphics[width=3.4in, clip, trim= 0cm 0.9cm 0cm 0cm]{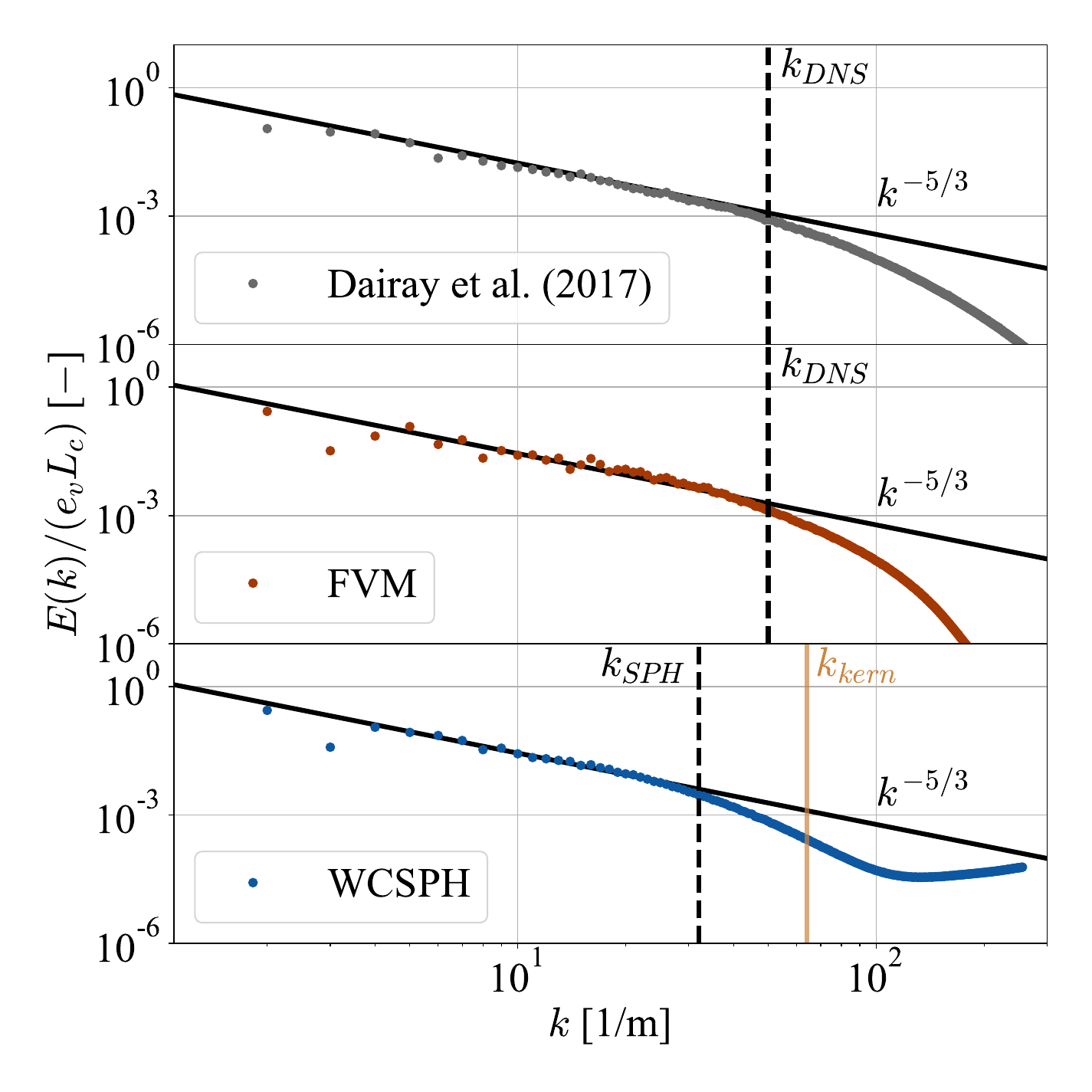}
\caption{\textcolor{black}{Comparison of kinetic energy spectra at $t=14~\mathrm{s}$. DNS of Dairay et al. \cite{Dairay_2017} is depicted in grey, the FVM LES in red and the WCSPH with the highest resolution in blue.}}
\label{fig:06_Spectra}
\end{figure}

Having verified that the global dynamics of the kinetic energy is reasonably well approximated by the WCSPH scheme, it remains to clarify whether HIT with the characteristic inertial range scaling in Eq. (\ref{eq:HIT_Energy_Scaling}) develops after the dissipation peak at $t \approx 9~\mathrm{s}$. Therefore, we consider the normalized kinetic energy spectra as explained in Sec. \ref{sec:HIT} for $t=14~\mathrm{s}$, which is in accordance with the work of Dairay \emph{et al.} \cite{Dairay_2017}. From their work an inertial range scaling should prevail up to a wavenumber of $k_{DNS} \approx 50~\mathrm{1/m}$. The computed spectra (TABLE \ref{tab:Cases}: Case 1-4) are visualized in FIG. \ref{fig:05_EnergyBalance} (e). For better orientation, the latter also contains a solid black line representing the $k^{-5/3}$ scaling and a dashed black line with a stronger $k^{-4}$ scaling. Moreover, the diagram includes the integral scale of the HIT problem (fawn dashed line) and the kernel scale of the simulation run with the highest particle count $N$ (fawn solid line). The corresponding $SNR$ values (Eq. (\ref{eq:SNR})) are also listed. Indeed it can be observed that the utilized WCSPH scheme without SFS model is able to recover a significant amount of the inertial range scaling for increasing $N$. While for the lowest $N$ the scaling is only evident in the range $k \in [4;~7]~\mathrm{1/m}$, the scaling range for the highest $N$ covers nearly an order of magnitude in wavenumber, namely $k \in [4;~32]~\mathrm{1/m}$. Hence, we can infer that WCSPH without SFS model is generally able to capture subsonic HIT, though at significant expense of $N=512^2$ and $N_{ngb} \approx 250$. This is consistent with the convergence properties described by Zhu \emph{et al.} \cite{Zhu_2015}. \textcolor{black}{The artificial thermalization as an anticipated origin of numerical dissipation is simultaneously reduced with increasing $N$, which is confirmed by an increase of the $SNR$ values in FIG. \ref{fig:05_EnergyBalance} (e). Nevertheless, even for the highest resolution, the artificial thermalization persists and likewise an approximate $k^{-4}$ scaling observable in all simulations that sets in well above the kernel scale. The latter connects the resolved inertial ranges with the artificial thermal ranges and represents a SPH characteristic kinetic energy deficit also observed by other authors, e.g. in Ref. \cite{Bauer_2012, Rennehen_2021}. It clearly demonstrates that the WCSPH method produces coarse-grained LES solutions from the beginning and in accordance with the 1. \underline{Implication} in Sec. \ref{subsec:DiscretizedLES}. However, the explicit kernel fails to realize a sharp cutoff close to $k_{kern}$, which could be due to the issue of \emph{Particle Duality} resulting in a decreased effective resolution. This is especially evident in FIG. \ref{fig:06_Spectra}, in which the spectra of the DNS, the FVM LES and the WCSPH simulation with the highest resolution (TABLE \ref{tab:Cases}: Case 4) are compared. Whereas the inertial range characteristics of the DNS (grey) and the FVM LES (red) are in perfect agreement and deviations are only evident beyond $k_{DNS} \approx 50~\mathrm{1/m}$, the WCSPH (blue) solution fails to reproduce the inertial range up to $k_{DNS}$. The spectrum is already damped at $k_{SPH}=32~\mathrm{1/m} = k_{kern}/2$, although $k_{kern} = 64~\mathrm{1/m}$ exceeds the maximum inertial range wavenumber $k_{DNS}$ of the reference DNS. Compared to the FVM LES with only $384^3$ cells, versus $N=512^3$ particles and $N_{ngb} \approx 250$ neighbors, this result is rather disappointing.}

\textcolor{black}{With these observations}, the aim of the next section is to investigate whether the explicit consideration of a dissipative SFS model can reduce the thermalization in \textcolor{black}{favor of scales larger than the kernel}. This should not only lead to an improvement \textcolor{black}{of the inertial range approximation} but also in terms of the global dynamics. Consequently, we will seek for evidence for the 2. \underline{Implication} in Sec. \ref{subsec:DiscretizedLES} following the argumentation line of Sec. \ref{sec:SFS}. We start with the \texttt{SMAG} model.

\begin{figure*}[t]
\includegraphics[width=7.0in, clip, trim= 0cm 2.6cm 0cm 2cm]{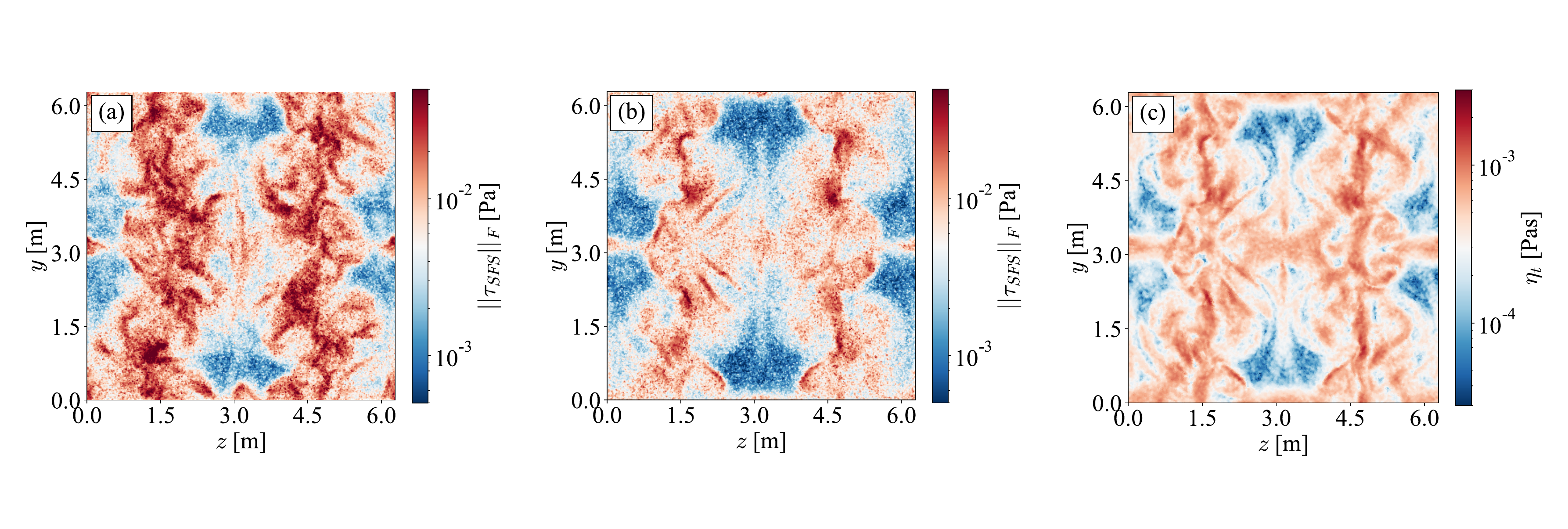}
\caption{(a) Frobenius norm of the estimated SFS tensor without explicit SFS model and (b) with \texttt{SMAG} model. (c) Turbulent dynamic eddy viscosity resulting from the \texttt{SMAG} model. All distributions refer to the $x=\pi$ plane at $t=14~\mathrm{s}$ for $N=256^3$.}
\label{fig:07_SFS_t14s_Comparison}
\end{figure*}

\subsection{\label{subsec:SmagorinskyEffect} Effect of the Smagorinsky Model}

Although the usage of a consistent LES model with an explicit SFS model should lead to an overall improvement of the solution, the investigation with the static \texttt{SMAG} model undermines this positive expectation (TABLE \ref{tab:Cases}: Case 5-8). Especially in terms of the global dynamics, represented by FIG. \ref{fig:05_EnergyBalance} (b) \& FIG. \ref{fig:05_EnergyBalance} (d), the consideration of the \texttt{SMAG} model only deteriorates the kinetic energy balance. This becomes evident from the temporal evolution of the averaged kinetic energy in FIG. \ref{fig:05_EnergyBalance} (b). For the whole time range the level of the averaged kinetic energy is reduced compared to the reference WCSPH solution in FIG. \ref{fig:05_EnergyBalance} (a). It holds true for all particle counts $N$ analyzed. The deterioration of the solution must also be apparent in the corresponding dissipation rates in FIG. \ref{fig:05_EnergyBalance} (d), which is indeed the case. A comparison of the WCSPH reference cases in FIG. \ref{fig:05_EnergyBalance} (c) with the \texttt{SMAG} cases in FIG. \ref{fig:05_EnergyBalance} (d) demonstrates in particular in the initial timeframe ($t < 7.5~\mathrm{s}$), up to the point where the blue dissipation lines cross the black DNS line, that the excessive dissipation of the WCSPH solution is significantly enhanced by the \texttt{SMAG} model. For the lowest particle count of $N=128^3$ it even results in a noticeable qualitative shift of the first dissipation peak from $t \approx 9~\mathrm{s}$ to $t \approx 6~\mathrm{s}$. For the remaining particle counts the position of the first dissipation peak in FIG. \ref{fig:05_EnergyBalance} (d) is quite robust, though a slight shift towards earlier times is perceptible. It is interesting to note that in this initial timeframe the vortex system is still in transition to HIT \cite{Brachet_1983, Dairay_2017, Pereira_2021}, which prompts the eventuality that this deterioration of the solution might be linked to the drawbacks of the Smagorinsky model as explained in Refs. \cite{Silvis_2017, Moser_2021}. In order to refute this eventuality, we will also present in Sec. \ref{subsec:OtherSFSModels} an investigation employing the superior \texttt{SIGMA} model (Appendix \ref{subsec:SFS}) .

From the discouraging global kinetic energy balance one might be tempted to conclude that the described link between explicit LES and SPH in Sec. \ref{sec:SPH} might be flawed. However, the kinetic energy spectra at $t=14~\mathrm{s}$ in FIG. \ref{fig:05_EnergyBalance} (f) reveal that the \textcolor{black}{anticipated causalities} presented in Sec. \ref{sec:SPH} are correct. Most importantly the comparison of the spectra in FIG. \ref{fig:05_EnergyBalance} (e) \& FIG. \ref{fig:05_EnergyBalance} (f) demonstrates the reduction of the artificial thermalization for a specific $N$, \textcolor{black}{although not very effectively}. The relative energy content below the kernel scales of the individual cases is mitigated by the \texttt{SMAG} model, which is consistent with a dissipative SFS model. This is also confirmed by an increase of the SNR metric (Eq. (\ref{eq:SNR})) for a given $N$, which are additionally listed in FIG. \ref{fig:05_EnergyBalance} (e) \& FIG. \ref{fig:05_EnergyBalance} (f). Moreover, this reduction of the thermalization correlates with a reduction of the Frobenius norm of $|| \boldsymbol{\tau}_{SFS} ||_F$, as anticipated in Sec. \ref{sec:SFS} and illustrated in FIG. \ref{fig:07_SFS_t14s_Comparison}. There, the estimated $||\boldsymbol{\tau}_{SFS} ||_F$ distribution is depicted for $N=256^3$ at the plane $x=\pi$ for $t=14~\mathrm{s}$, corresponding to the time instance of the energy spectra. Without explicit SFS (FIG. \ref{fig:07_SFS_t14s_Comparison} (a)) and with \texttt{SMAG} model (FIG. \ref{fig:07_SFS_t14s_Comparison} (b)) high values of $||\boldsymbol{\tau}_{SFS} ||_F$ prevail in the shear zones of the flow (cp. FIG. \ref{fig:04_InitialHIT}). However, as a consequence of the \texttt{SMAG} model, the absolute SFS values are finally reduced. This damping is associated to the turbulent dynamic eddy viscosity $\eta_t$ distribution (FIG. \ref{fig:07_SFS_t14s_Comparison} (c)), which evidently correlates with the $|| \boldsymbol{\tau}_{SFS} ||_F$ distribution (FIG. \ref{fig:07_SFS_t14s_Comparison} (a) \& FIG. \ref{fig:07_SFS_t14s_Comparison} (b)). By construction of Eq. (\ref{eq:SPH_SFS_10}), this is plausible as the turbulent eddy viscosity reacts to high values of spatial velocity gradients prevailing in the shear flow planes.
\begin{figure*}[ht]
\includegraphics[width=6.2in, clip, trim= 0cm 3.2cm 0cm 2cm]{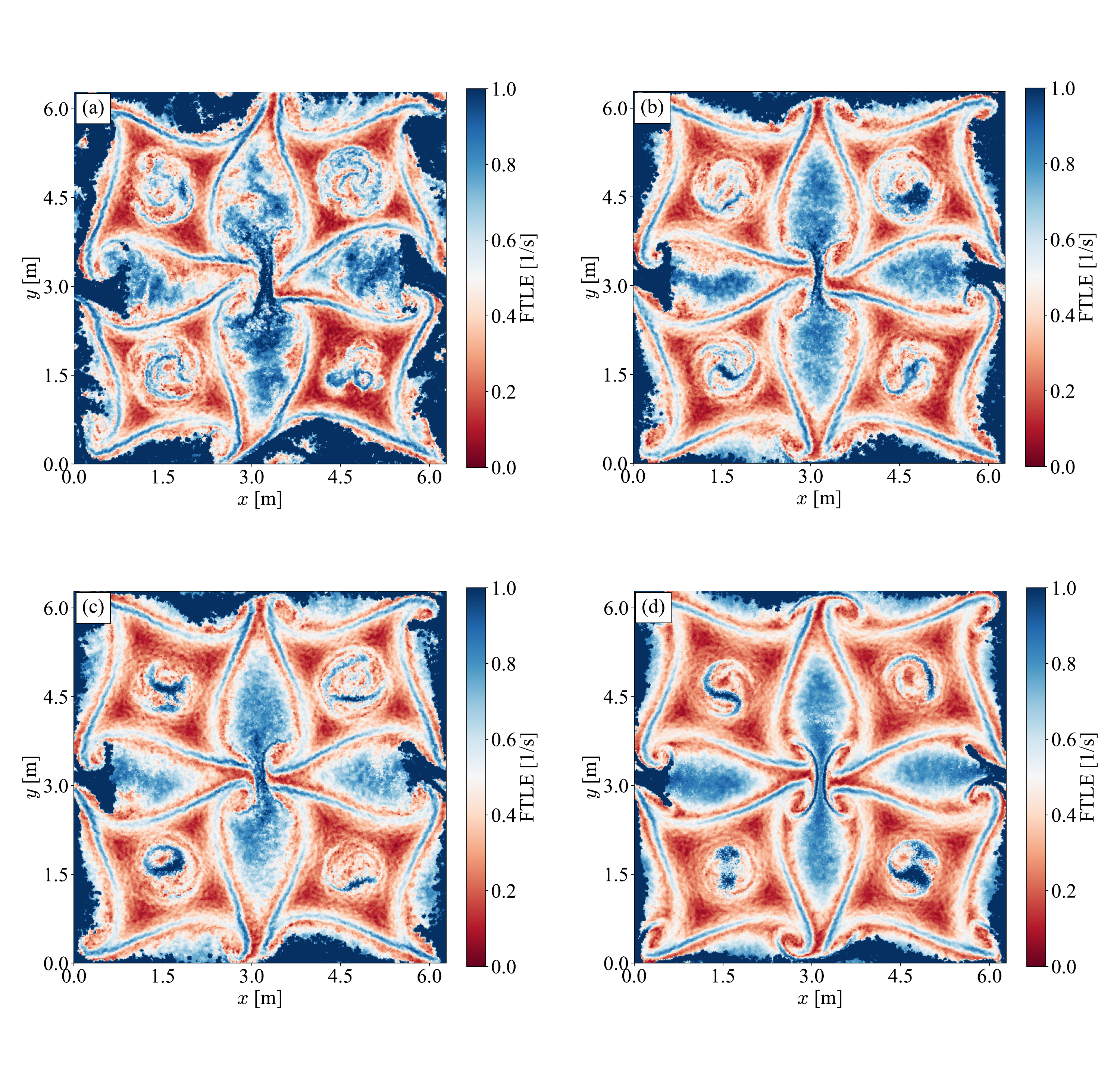}
\caption{Backward FTLE at the plane $z=\pi$ for $N=256^3$ and $t=14~\mathrm{s}$ in the range $[11; 14]~\mathrm{s}$. (a) Without explicit SFS model. (b) With \texttt{SMAG} model. (c) With \texttt{SIGMA} model. (d) With \texttt{SMAG-MCG} model.}
\label{fig:08_FTLE_Comparison}
\end{figure*}

For the remainder of this section we will discuss whether \textcolor{black}{scales larger than the kernel can} profit from the mitigated artificial thermalization. The answer to this is ambivalent and we start with the positive aspects. As a matter of fact, the relative energy content of the first wavenumber shells ($k \leq 7~\mathrm{1/m}$) increases as the comparison of FIG. \ref{fig:05_EnergyBalance} (e) \& \ref{fig:05_EnergyBalance} (f) demonstrates. Although the differences might seem extraneous, we want to emphasize that the plots are double logarithmic. Interestingly, this improved spectral signature can also be linked to physical space \textcolor{black}{for particle counts $N \leq 256^3$} by means of the backward finite-time Lyapunov exponent (FTLE) in the time range $[11, 14]~\mathrm{s}$. Slices of the resulting FTLE fields at the plane $z=\pi$ from the $N=256^3$ runs without explicit SFS model and with \texttt{SMAG} model are illustrated in FIG. \ref{fig:08_FTLE_Comparison} (a) \& FIG. \ref{fig:08_FTLE_Comparison} (b). Apparently, the resulting fields are representative for the coherent large scale vortices remaining from the initialization. Generally, the FTLE fields are quite similar in their appearance, however, it is undeniable that the structures formed in the cases with the \texttt{SMAG} model in FIG. \ref{fig:08_FTLE_Comparison} (b) are less tattered than the reference WCSPH solution in FIG. \ref{fig:08_FTLE_Comparison} (a). The most positive difference between the fields is that the consideration of the \texttt{SMAG} model approximately restores the mirror symmetry of the vortex system at the midplanes $x=\pi$ \& $y=\pi$. This mirror symmetry is a characteristic of the vortex systems \cite{Brachet_1983} and a vivid prove that a reduction of the artifical thermalization can positively influence the large scale coherent motion. As a sidenote, symmetry breaking of a similar kind was also observed in molecular approximations of the Taylor-Green system in the work of Gallis \emph{et al.} \cite{Gallis_2021}, albeit caused by physical thermal fluctuations. 

\begin{figure*}[t]
\includegraphics[width=7.0in, clip, trim= 0cm 2.6cm 0cm 2cm]{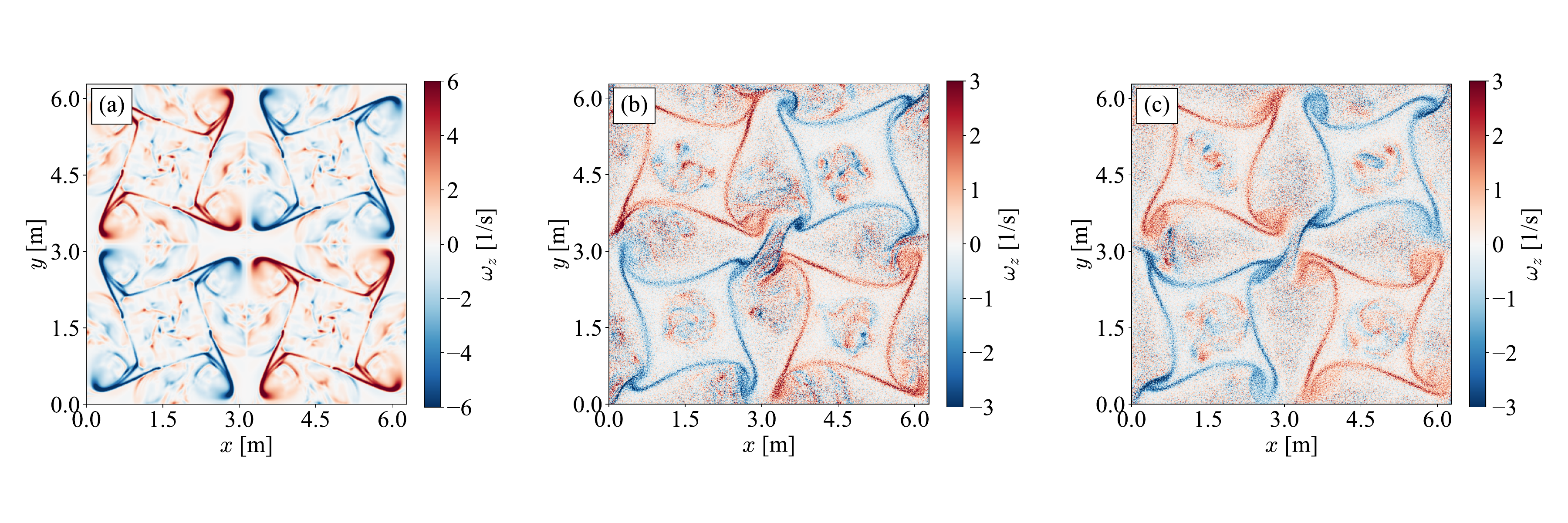}
\caption{\textcolor{black}{Vorticity component $\omega_z$ at the plane $z=\pi$ \& $t=14~\mathrm{s}$. (a) FVM LES with $384^3$ cells. (b) WCSPH with $N=512^3$ particles without explicit SFS. (c) SPH with $N=512^3$ particles and \texttt{SMAG} model.}}
\label{fig:09_Omega_Comparison}
\end{figure*}

\textcolor{black}{Unfortunately, it must also be accentuated that the symmetry restoration is not discernible anymore for the highest resolution SPH runs. This is depicted in FIG. \ref{fig:09_Omega_Comparison} (b) \& (c). There, the vorticity field component $\omega_z$ at $z=\pi$ \& $t=14~\mathrm{s}$ of the highest resolution SPH runs without and with \texttt{SMAG} model (TABLE \ref{tab:Cases}: Case 4 \& 8) are illustrated. Obviously, the quantity behaves very similar to the FTLE field in FIG. \ref{fig:08_FTLE_Comparison}. The lost symmetry restoration property is quite disadvantageous given that the $N=512^3$ cases are the only runs, which reproduce a significant amount of inertial range scaling. Although the qualitative large scale pattern is matched in comparison with the symmetric FVM field in FIG. \ref{fig:09_Omega_Comparison} (a), the SPH solutions evidently tend to smear out sharp features of the large scale vortices. This accordingly results in smaller vorticity values as the colorbars in FIG. \ref{fig:09_Omega_Comparison} illustrate.}

\textcolor{black}{Despite the slight improvements described before, the overall effect of the \texttt{SMAG} model is rather malicious. The deterioration of the global dynamics in terms of the averaged kinetic energy in FIG. \ref{fig:05_EnergyBalance} (b) and the corresponding dissipation rate in FIG. \ref{fig:05_EnergyBalance} (d) strongly indicate that the negative aspects outweigh the positive ones. This is also reflected in the energy spectra in FIG. \ref{fig:05_EnergyBalance} (f) for $k > 7~\mathrm{1/m}$ and $k<k_{kern}$. A comparison of FIG.\ref{fig:05_EnergyBalance} (e) \& \ref{fig:05_EnergyBalance} (f) clearly shows that the effect of the dissipative \texttt{SMAG} model is not restricted to scales below the kernel. Not only the energy of the artificial thermalization is reduced, but also the kinetic energy deficit above the kernel scale. Hence, the deficient $k^{-4}$ scaling of SPH is expanded to a larger wavenumber range at the expense of the reproduced inertial range. This proves that the \emph{Particle Duality} described in Sec. \ref{subsec:ParticleDuality} truly prevents the dissipative \texttt{SMAG} model to operate solely on noisy scales smaller than the kernel, which trigger numerical dissipation. Consequently, from these observations, the application of the \texttt{SMAG} model in SPH should be discouraged. Nevertheless, one might infer that these results are the consequence of a wrongly calibrated  \texttt{SMAG} model or the \texttt{SMAG} model itself, instead of a consequence of the \emph{Particle Duality} of SPH. Thus, we will discuss the influence of alternative SFS models and the calibration of the \texttt{SMAG} model in the next section.}

\subsection{\label{subsec:OtherSFSModels} Other Dissipative SFS Models and Model Calibration}

\begin{figure*}
\includegraphics[width=5.7in, clip, trim= 0cm 0cm 0cm 0cm]{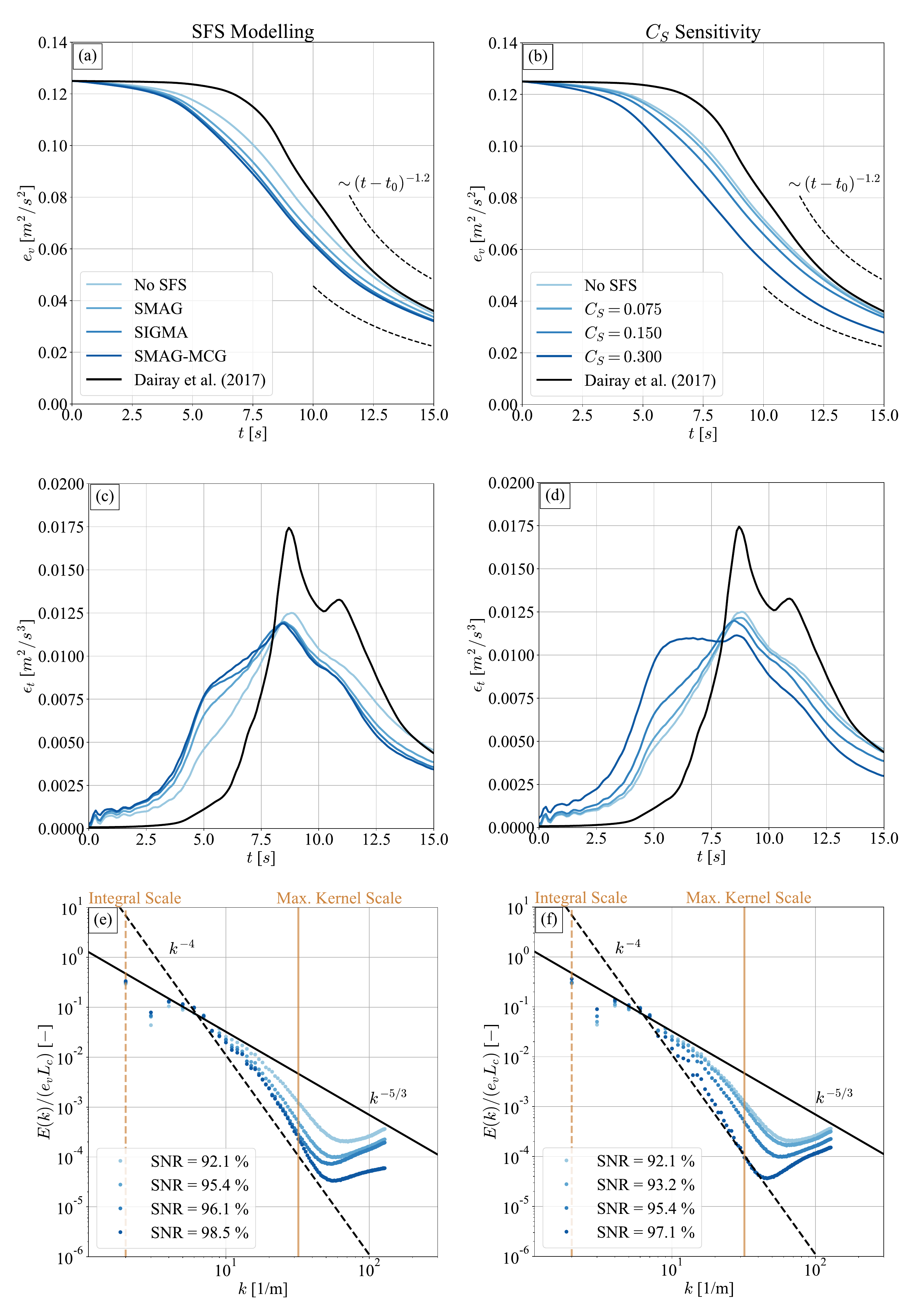}
\caption{\textcolor{black}{Comparison of quantitative metrics for different $N=256^3$ runs. The first column (a), (c) \& (e) represents the influence of different explicit SFS models and the second column (b), (d) \& (f) represents a sensitivity study of the Smagorinsky constant $C_S$ for the \texttt{SMAG} model. The color coding is explained either in (a) or (b). (a) \& (b) Temporal evolution of the density
weighted averaged kinetic energy $e_v$. (c) \& (d) Temporal evolution of the averaged dissipation rate $\epsilon_t$. (e) \& (f) Kinetic energy spectra at $t=14~\mathrm{s}$.}}
\label{fig:10_EnergyBalance_SFS}
\end{figure*}

In this part a comparison between the WCSPH solution without SFS model and with \texttt{SMAG}, \texttt{SIGMA} \& \texttt{SMAG-MCG} model will be presented, \textcolor{black}{as well as a sensitivity study of the Smagorinsky constant $C_S$} (TABLE \ref{tab:Cases}: Case 3 \& 9-12). As the observations seem to be independent from the particle count $N$, only the $N=256^3$ runs will compared. The results are depicted in FIG. \ref{fig:10_EnergyBalance_SFS}.  All in all, the observations are very similar to those presented in Sec. \ref{subsec:SmagorinskyEffect}. \textcolor{black}{We will start with the influence of the explicit SFS models (TABLE \ref{tab:Cases}: Case 3, 7 \& 9-10).} 

Neither the superior \texttt{SIGMA} model nor the discrete angular momentum conserving \texttt{SMAG-MCG} model \textcolor{black}{lead to an improvements in terms of the kinetic energy characteristics.} In fact, both models are even worse. As depicted in FIG. \ref{fig:10_EnergyBalance_SFS} (a), the averaged kinetic energy levels of both variants are slightly below the \texttt{SMAG} solution. The global dissipation rates in FIG. \ref{fig:10_EnergyBalance_SFS} (c) confirm these results. Accordingly, this is also reflected by the spectra in FIG. \ref{fig:10_EnergyBalance_SFS} (e). This is surprising for different reasons. For the \texttt{SIGMA} model a vanishing dissipation in the initial laminar phase by the nature of the SFS model \cite{Nicoud_2011, Silvis_2017} would be expected. However, as depicted in FIG. \ref{fig:10_EnergyBalance_SFS} (c), the overall dissipation is increased from the beginning. One might be tempted to concluded that this is related to the missing zero order consistency of the SPH-LES model in Eq. (\ref{eq:LES_Quadrature}), prohibiting the flow discrimination required for the \texttt{SIGMA} model. Nonetheless, similar observation were made using high-order Eulerian grid based schemes utilizing the same SFS model for Taylor-Green flows \cite{Fernandez_2017, Evans_2018}. Consequently, this indicates that the Taylor-Green flow is a challenging problem for the \texttt{SIGMA} model independent of the numerical discretization scheme. The kinetic energy spectrum in FIG. \ref{fig:10_EnergyBalance_SFS} (e) further demonstrates that the qualitative effect of the \texttt{SIGMA} model and the \texttt{SMAG} model are similar. Compared to the \texttt{SMAG} model, the artificial thermalization below the kernel scale is slightly reduced, from which the first wavenumber shells ($k \leq 7~\mathrm{1/m}$) \textcolor{black}{again slightly profit}. This is also reflected by the FTLE field in FIG. \ref{fig:08_FTLE_Comparison} (c), which compared to the WCSPH solution in FIG. \ref{fig:08_FTLE_Comparison} (a) is less tattered and, moreover, approximately mirror symmetric at the midplanes. \textcolor{black}{However and most importantly,} the dissipative \texttt{SIGMA} model still suffers from the issue of \emph{Particle Duality}. \textcolor{black}{A removal of kinetic energy from scales in the range $k > 7~\mathrm{1/m}$ and $k<k_{kern}$ is still present and even leads to an intensification of the observed deficient SPH energy scaling with $E(k)\sim k^{-n},~n>4$.} The \texttt{SMAG-MCG} model results \textcolor{black}{tell a nearly identical story}. Compared to the \texttt{SMAG} model the overall dissipation is enhanced. This becomes evident in FIG. \ref{fig:10_EnergyBalance_SFS} (a) \& FIG. \ref{fig:10_EnergyBalance_SFS} (c). It is surprising as one would intuitively expect a general improvement related to the restoration of the angular momentum conservation property. Instead, the considered problem demonstrates that this comes at a certain cost. The kinetic energy spectrum in FIG. \ref{fig:10_EnergyBalance_SFS} (e) once more confirms the already noted observations. The \texttt{SMAG-MCG} model is characterized by the strongest reduction of the artificial thermalization, which repeatedly has a positive effect on the first wavenumber shells ($k \leq 7~\mathrm{1/m}$) in the spectra, as well as the FTLE field in FIG. \ref{fig:08_FTLE_Comparison} (d). Nevertheless, the issue of \emph{Particle Duality} for the \texttt{SMAG-MCG} model in the range $k > 7~\mathrm{1/m}$ and $k<k_{kern}$ is yet evident. Compared to the \texttt{SMAG} model, the angular momentum conservation property of the MCG form intensifies the observed deficient SPH scaling $E(k)\sim k^{-n},~n>4$.

\textcolor{black}{So far our results can be criticized in terms of the fact that the utilized dissipative SFS models might be wrongly calibrated for the considered problem and discretization method \cite{Moser_2021}. For instance in the Smagorinsky model in Eq. (\ref{eq:SPH_SFS_10}), one has the freedom to choose the constant $C_S$ and the filter width $\Delta$. Since we set $\Delta=D_K/2$ in Sec. \ref{sec:HIT}, subsequently the influence of $C_S$ is analyzed (TABLE \ref{tab:Cases}: Case 11-12). Compared to our reference, the value of $C_S$ is either halved or doubled. We want to note that this is identical to the situation in which $C_S=0.15$ is fixed and $\Delta$ changed accordingly. The results are depicted in the second column of FIG. \ref{fig:10_EnergyBalance_SFS} and the increase of $C_S$ (or $\Delta$) results in a distinct monotonous behaviour. The averaged kinetic energy is increasingly reduced within the considered time range as shown in FIG. \ref{fig:10_EnergyBalance_SFS} (b), which is also logically reflected by the dissipation rates in FIG. \ref{fig:10_EnergyBalance_SFS} (d). For the highest value $C_S=0.03$, even an artificial dissipation rate plateau within the range $t \in [6;~9]~\mathrm{s}$ is created. The dynamics increasingly departs from the case without SFS model for larger $C_S$ (or $\Delta$). Same holds for the energy spectra in FIG. \ref{fig:10_EnergyBalance_SFS} (f). It is interesting to note that the artificial thermalization is only slightly reduced compared to the range where the deficient energy scaling of $k^{-4}$ prevails. This shows again the non-local character induced by the issue of \emph{Particle Duality}. It becomes more pronounced with larger $C_S$ and incrementally represses the inertial range, showing that the best choice corresponds to $C_S \to 0$ (or $\Delta \to 0$).}

\textcolor{black}{Clearly, from all these observations, it can be concluded that WCSPH without explicit SFS generates coarse-grained solution from the beginning, however at significant cost. This is likely due to the \emph{Particle Duality} issue, which phenomenologically indicates that the method suffers from an decreased effective resolution. The intrinsic property also seems to prohibit improvements by means of standard dissipative SFS models. Although the models mitigate the artificial thermalization, which is believed to be the origin of numerical dissipation and the largest resolved scales can slightly profit from it, the SFS model dominantly remove kinetic energy from scales larger than the kernel. These scales are already badly resolved in terms of the energy cascade, which let us confirm the statement of Rennehen that dissipative SFS models overall degrade the SPH solution \cite{Rennehen_2021}.}

\textcolor{black}{Having verified that the best SPH-LES model relinquishes the explicit usage of standard dissipative SFS models, we finally proceed with an investigation of numerical dispersion effects for the pure WCSPH scheme.}

\subsection{\label{subsec:Dispersion}Numerical Dispersion and Galilean Invariance}

\textcolor{black}{In this short paragraph, we will discuss whether numerical dispersion errors in the WCSPH scheme (Eq. (\ref{eq:LES_Quadrature})) might influence the turbulent dynamics. Therefore, based on the idea of Yalla et al. \cite{Yalla_2021}, we investigate whether our overall model is Galilean invariant. If the energy cascade is significantly altered, this could indicate that dispersion errors inhibit the transfer to the smallest scales.
}
\begin{figure}[h]
\includegraphics[width=3.in, clip, trim= 0cm 0.9cm 0cm 0cm]{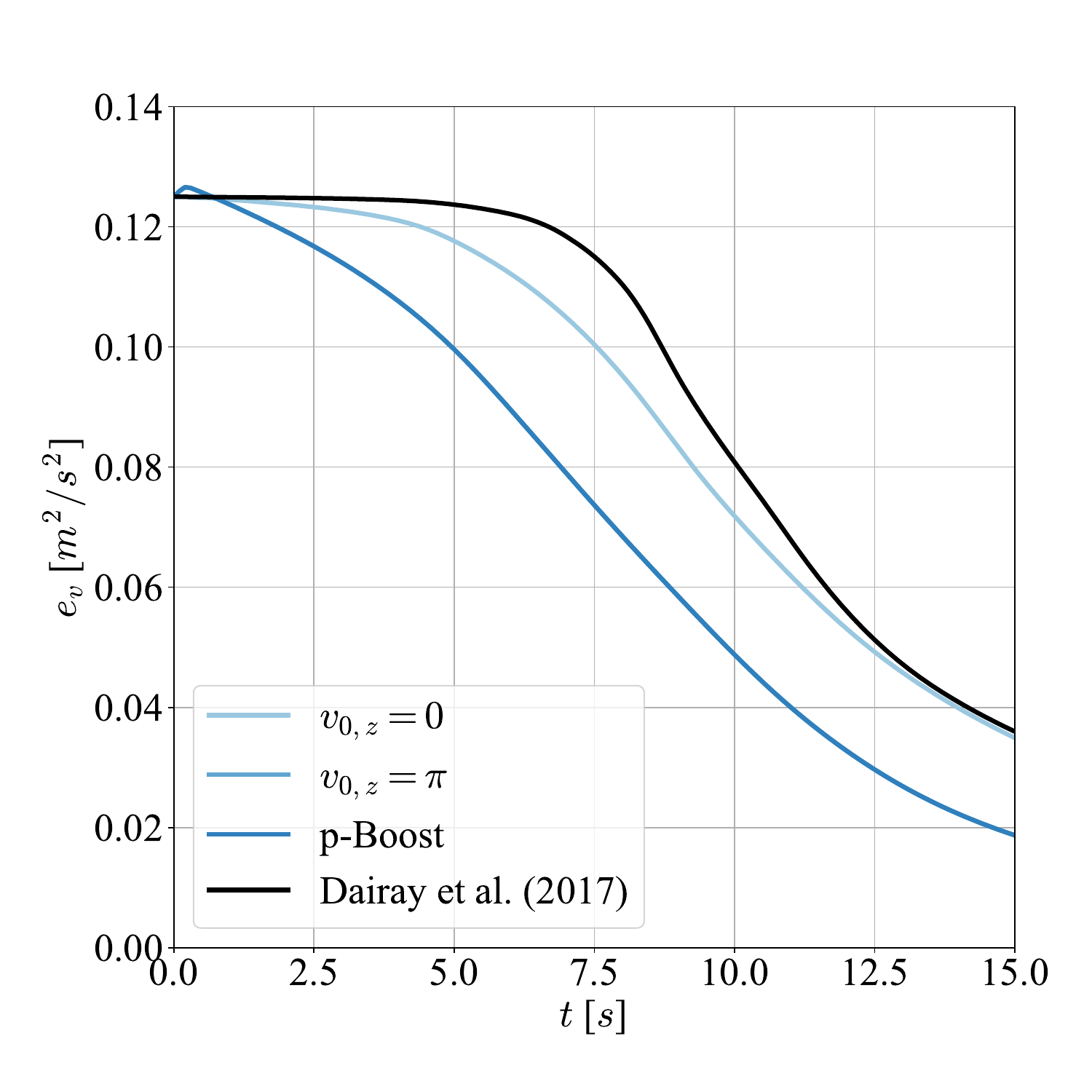}
\caption{\textcolor{black}{Temporal evolution of the density
weighted averaged kinetic energy $e_v$ obtained from the analysis of Galilean invariance. The dissipation of the convected case with $v_{0,z}=\pi$ is strongly increased, however, this is only the consequence of the increase of $p_{ref}$ as the case p-Boost with $v_{0,z}=0$ demonstrates. The lines for $v_{0,z}=\pi$ and p-Boost perfectly overlap each other.}}
\label{fig:11_KineticEnergy_pBoost}
\end{figure}

\begin{figure}[h]
\includegraphics[width=3.in, clip, trim= 0cm 0.9cm 0cm 0cm]{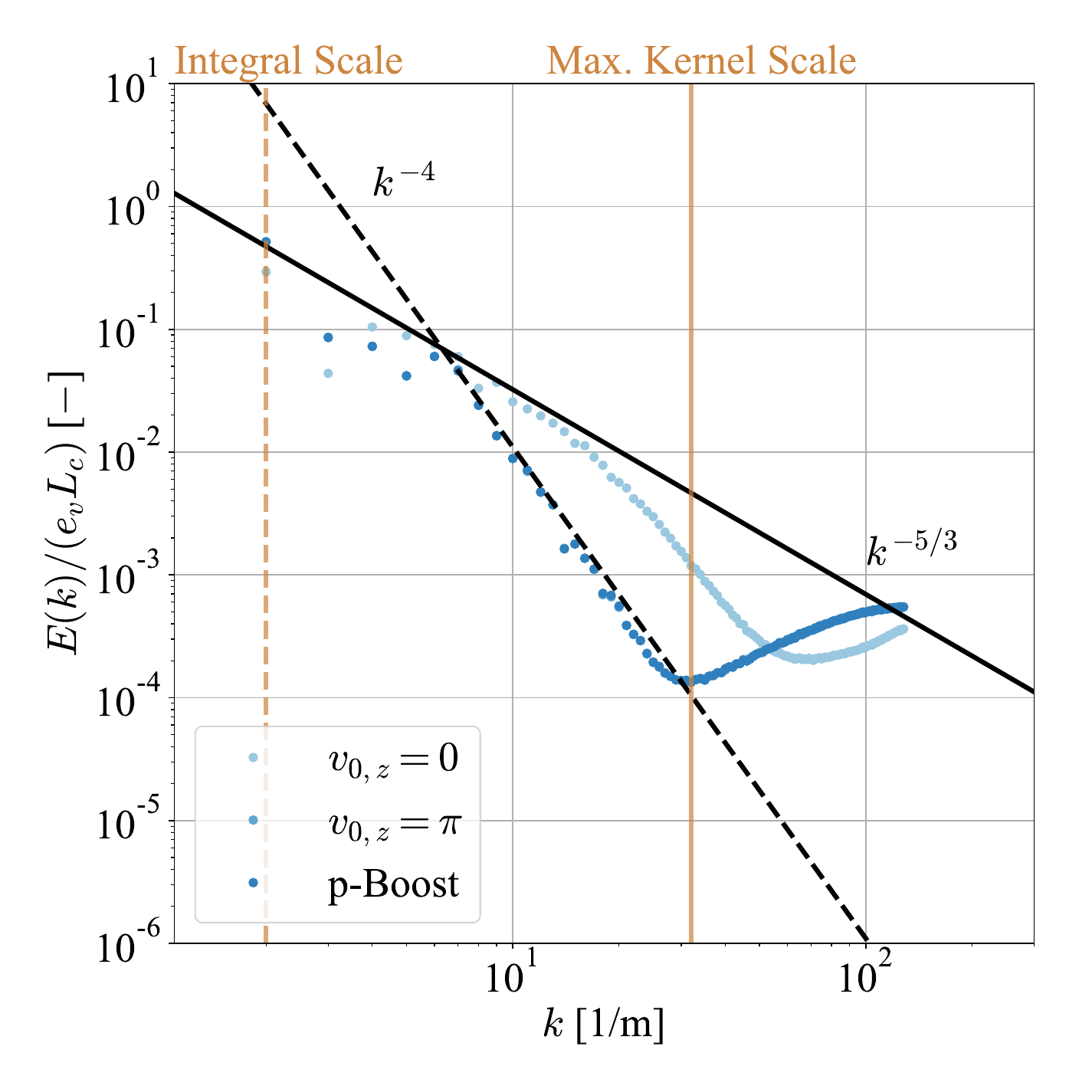}
\caption{\textcolor{black}{Kinetic energy spectra at $t=14~\mathrm{s}$ obtained from the analysis of Galilean invariance. The spectrum of the convected case with $v_{0,z}=\pi$ is strongly altered, however, this is only the consequence of the increase of $p_{ref}$ as the case p-Boost with $v_{0,z}=0$ demonstrates. The lines for $v_{0,z}=\pi$ and p-Boost perfectly overlap each other.}}
\label{fig:12_Spectra_pBoost}
\end{figure}

\textcolor{black}{However, before we compare the convected Case 13 with $\langle v_z \rangle_V = \pi$ and the quiescent Case 3, we will first study how Case 3 is influenced by the increase of $p_{ref}$, which results from the requirement to keep initial Mach similarity (see Sec \ref{sec:HIT}). The case is termed as "p-Boost". Evidently, as illustrated in FIG. \ref{fig:11_KineticEnergy_pBoost}, the global dissipation is heavily increased. Even a short initial phase is created in which kinetic energy is injected into the flow. This could be the consequence of a weak compressibilty effect. The overall increase of dissipation is comprehensible, considering that the value $p_{ref}$ is adjusting the magnitude of a force, which is the consequence of zero order errors only \cite{Price_2012,Colagrossi_2012, Read_2010}. Its corresponding acceleration reads}
\begin{equation}
    \mathbf{a}_i = \frac{2p_{ref}}{\rho_i}\sum_{j=1}^{N_{ngb}}\nabla W_{h,ij}V_j~
    \label{eq:pref_ZeroOrder}
\end{equation}
\textcolor{black}{and physically causes an artificial momentum transport of particles that avoids the creation of void spaces. It ensures the stability of the method due to an inherent particle regularization, which reacts to irregular distributions, where the ideal condition $\sum_{j=1}^{N_{ngb}}\nabla W_{h,ij}V_j = 0$ is violated. It can be anticipated that this artificial transport will also take place perpendicular to the local main flow directions, causing numerical dissipation according to the ideas presented in Sec. \ref{sec:SFS}. However, from the generic SPH dispersion study of Dehnen \& Aly \cite{Dehnen_2012}, it is certain that not only numerical dissipation with $p_{ref}$ is increased but also the influence of dispersion errors. This is reflected in the spectra in FIG. \ref{fig:12_Spectra_pBoost}. Beyond the increase of the artificial thermalization, which likely reflects the increase of numerical dissipation, the inertial range is even stronger suppressed in the p-Boost case. The deficient SPH scaling $k^{-4}$ profits from increased $p_{ref}$ value. This could be an indication that dispersion errors inhibit the transfer of energy to the smallest scales, spoiling the development of an appropriate inertial range similiar to the study of Yalla et al. \cite{Yalla_2021}.  }

\textcolor{black}{Ensuing from the p-Boost case, the results of the convected Case 13 with $\langle v_z \rangle_V = \pi$ are comprehensible. As depicted in FIG. \ref{fig:11_KineticEnergy_pBoost} \& \ref{fig:12_Spectra_pBoost}, the lines for $v_{0,z}=\pi$ and p-Boost perfectly overlap each other. This shows that the WCSPH scheme is truly Galilean invariant, though only if the mean convective velocity is changed. Practically, the superposition of a mean flow also requires an adjustment of $p_{ref}$ to ensure stability, which evidently will spoil the Galilean invariance. The zero order term in Eq. (\ref{eq:pref_ZeroOrder}) can drastically increase numerical dissipation and also inhibit the development of an inertial range due to the increase of dispersion errors. Consequently, the results obtained in our work might be rather positive compared to cases in which a mean convection is superimposed onto the turbulent flow. It is certain that this interplay of numerical dissipation and dispersion caused by  $p_{ref}$ requires a more thorough study, but from the selected tests analyzed in this section the first observations seem reasonable.}

\section{\label{sec:Conclusion} Conclusion}

Summarizing, the contributions of this work are numerous and especially important for WCSPH simulations trying to capture subsonic turbulent flows. Our main goal was to argue in Sec. \ref{sec:Turbulence} and Sec. \ref{sec:SPH} that, based on the Hardy theory from NEMD, SPH should be viewed as a non-local Lagrangian quadrature procedure intrinsically related to explicit LES. This implies on the one hand that subsonic turbulence captured by SPH will be correctly represented, in the best case, up to the kernel scale but also at significant cost, taking the convergence characteristics into account \cite{Zhu_2015, Sigalotti_2021}. This is consistent with observations in the literature \cite{Bauer_2012, Zhu_2015}, empirically supporting our LES perspective on SPH. On the other hand, it paves a potential way to mitigate SPH characteristic shortcomings by explicit consideration of a dissipative SFS model as explained in Sec. \ref{sec:SFS}. \textcolor{black}{The main idea followed herein was to replace the numerical dissipation, emerging from the artificial thermalization, by an explicit, dissipative SFS model.} In order to test the hypothesis that a reduction of the artificial thermal range results in an improvement of the kinetic energy content of \textcolor{black}{scales larger than the kernel}, several simulations of freely decaying HIT at $Re=10^4$ in accordance to Dairay \emph{et al.} \cite{Dairay_2017} were conducted and analyzed. However, it must be stated from the results presented in Sec. \ref{sec:Results} that the explicit SFS model only leads to an \textcolor{black}{marginal} improvement of the largest coherent structures. This was vividly reflected by the symmetry restoration of the vortex system in the computed FTLE fields. \textcolor{black}{Concerning the inertial range dynamics}, the dissipative SFS models merely remove kinetic energy where SPH is already characterized by a spectral energy deficit. Eventually, it deteriorates the overall solution outweighing the positive effects. This is rooted in the non-local character of the Lagrangian quadrature, which can be explained by the concept of \emph{Particle Duality}. The latter states that the SPH particles must simultaneously represent super fluid element approximants and fluid element surrogates at the same time, causing an non-physical increase of the effective particle interaction distance. Finally, our work allows to confirm Rennehen's expectation that explicit SFS models in a SPH framework only degrade the quality of the approximation for subsonic turbulent flows \cite{Rennehen_2021} and from our current understanding they should be disregarded. It seems to be the case that the \textcolor{black}{excessive numerical} dissipation mechanisms of SPH \cite{Ellero_2010, Colagrossi_2013, Okraschevski_2021_1} outperform the explicit dissipative SFS models considered in this work. 
\textcolor{black}{However, only adapted SFS models, originally developed for grid-based Eulerian methods, were tested. This possibly indicates that SPH native SFS models are needed that consider SPH specific characteristics. }

\textcolor{black}{Furthermore, our work obviously shows that the understanding of numerical dissipation and dispersion in kernel-based methods like SPH and their influence on turbulent dynamics is very heuristic compared to conventional Eulerian grid-based methods. This lacking rigor must be definitely addressed in future works.}

As next step it would be interesting to study whether the issue of \emph{Particle Duality} in terms of explicit dissipative SFS model can be either circumvented by higher-order schemes or by the use of the SPH-LES scheme developed by Di Mascio \emph{et al.}\cite{DiMascio_2017} and Antuono \emph{et al.}\cite{Antuono_2021}. Although we currently believe that the issue of \emph{Particel Duality} is a conceptual problem of SPH, there may be the chance that cross-effects between different modelling terms can restore the actual goal of standard explicit dissipative SFS models.

\begin{acknowledgments}
\textcolor{black}{The authors would like to thank Eric Lamballais for kindly sharing the energy spectrum of the DNS presented in Ref. \cite{Dairay_2017}}. Moreover, the authors acknowledge support by the state of Baden-Württemberg through bwHPC.
\end{acknowledgments}

\section*{Data Availability Statement}

The data that support the findings of this study are available from the corresponding author upon reasonable request.

\appendix

\section{\label{sec:Appendix} SPH Discretization of the Explicit LES model}

\subsection{\label{subsec:Density}The Averaged Density}

In order to approximate the averaged density of a specific particle $i \in \{1,~...,~N\}$ at position $\mathbf{x} = \mathbf{x}_i$, the integral expression in Eq. (\ref{eq:LES_Mass}) has to be discretized. Therefore, we define a mass differential for the Lagrangian element $\mathbf{y}$ with $\mathrm{d} M(\mathbf{y}) := \rho \mathrm{d}\mathbf{y}$, which allows one to rewrite Eq. (\ref{eq:LES_Mass}) into
\begin{equation}
    \overline{\rho}_i := \overline{\rho} (\mathbf{x}_i, t) = \int\displaylimits_{V_x}  W_h(\mathbf{x} - \mathbf{y})~\mathrm{d}M(\mathbf{y})~.
    \label{eq:DensityMassElement}
\end{equation}
Assuming that $V_x$ after the decomposition contains $j \in \{1,~...,~N_{ngb}\}$ neighbor (ngb) particles, a naive quadrature can be applied, in which $\mathrm{d}M(\mathbf{y})$ is replaced by a finite mass at position $\mathbf{y} = \mathbf{y}_j$, namely $M_j := M(\mathbf{y}_j)$. With the abbreviation $W_{h,ij}:=W_h(\mathbf{x}_i - \mathbf{y}_j)$, this gives
\begin{equation}
    \overline{\rho}_i = \sum_{j=1}^{N_{ngb}} M_j W_{h,ij} + \mathcal{O} (N_{ngb}^{-\gamma})~, \quad \gamma \in [\frac{1}{2}; 1]~.
    \label{eq:SPHDensity_1}
\end{equation}
The approximation in Eq. (\ref{eq:SPHDensity_1}) is the standard way of density estimation in SPH \cite{Monoghan_2005, Springel_2010, Price_2012} and contains an error term, which vanishes with $N_{ngb} \to \infty$ according to Ref. \cite{Sigalotti_2021,Zhu_2015}. 

For a homogeneous distribution of mass corresponding to $M_i=M_j$, Eq. (\ref{eq:SPHDensity_1}) implies a definition for the particle volume
\begin{equation}
    \overline{\rho}_i \approx M_i \sum_{j=1}^{N_{ngb}}  W_{h,ij} = \frac{M_i}{V_i} \quad \Rightarrow \quad V_i := \frac{1}{\sum_{j=1}^{N_{ngb}}  W_{h,ij}}~.
    \label{eq:SPHDensity_2}
\end{equation}
Moreover, Eq. (\ref{eq:SPHDensity_2}) represents another common way of density approximation in SPH, which is often used in subsonic multiphase problems with density discontinuity even when $M_i\neq M_j$ \cite{Espanol_2003, Hu_2006}. Hence, Eq. (\ref{eq:SPHDensity_2}) is chosen as density approximation in our work due to its higher level of generality, although we utilize an inhomogeneous mass distribution (see Sec. \ref{sec:HIT}). However, we tested both approximations (Eq. (\ref{eq:SPHDensity_1}) \& Eq. (\ref{eq:SPHDensity_2})) finding negligible influence on our results.

Contrary to the usual SPH approach, Eq. (\ref{eq:DensityMassElement}) is exact and should not be interpreted as a smoothed approximation of the true fluid density. Our goal is to approximate the density of LES super fluid elements, hence the Lagrangian quadrature applied in Eq. (\ref{eq:SPHDensity_1}) is the only approximation which is introduced.

\subsection{\label{subsec:Pressure}The Averaged Pressure Gradient}

For the averaged pressure gradient in the momentum balance of $V_x$ in Eq. (\ref{eq:LES_Momentum}), we will first apply integration by parts
\begin{eqnarray}
    &-& \int\displaylimits_{V_x} \nabla_\mathbf{y} p(\mathbf{y},t) W_h(\mathbf{x} -  \mathbf{y})~\mathrm{d}\mathbf{y}  \label{eq:SPHPressureGrad_1} \\
    = &-& \int\displaylimits_{V_x} \nabla_\mathbf{y} [ p(\mathbf{y},t) W_h(\mathbf{x} -  \mathbf{y}) ]~\mathrm{d}\mathbf{y}  
    + \int\displaylimits_{V_x}  p(\mathbf{y},t) \nabla_\mathbf{y} W_h(\mathbf{x} -  \mathbf{y}) ~\mathrm{d}\mathbf{y}~. \nonumber
\end{eqnarray}
Using the theorem of Gauss-Ostrogradsky with $\mathrm{d}\mathbf{o}$ as surface differential and the fact that $W_h = 0$ on the boundary $\partial V_x $ of the super fluid element $V_x$ by definition, the first term on the right hand side of Eq. (\ref{eq:SPHPressureGrad_1}) takes the form
\begin{equation}
     \int\displaylimits_{\partial V_x}  p(\mathbf{y},t) W_h(\mathbf{x} -  \mathbf{y}) ~\mathrm{d}\mathbf{o} = 0
    \label{eq:SPHPressureGrad_2}
\end{equation}
and vanishes. For the second term on the right hand side of Eq. (\ref{eq:SPHPressureGrad_1}) the chain rule can be used to demonstrate that $\nabla_\mathbf{y} W_h(\mathbf{x} -  \mathbf{y}) = - \nabla_\mathbf{x} W_h(\mathbf{x} -  \mathbf{y})$ and this results in the exact expression
\begin{eqnarray}
    &-& \int\displaylimits_{V_x} \nabla_\mathbf{y} p(\mathbf{y},t) W_h(\mathbf{x} -  \mathbf{y})~\mathrm{d}\mathbf{y} \label{eq:SPHPressureGrad_3} \\
    = &-& \int\displaylimits_{V_x}  p(\mathbf{y},t) \nabla_\mathbf{x} W_h(\mathbf{x} -  \mathbf{y}) ~\mathrm{d}\mathbf{y}~. \nonumber
\end{eqnarray}
Before we apply a Lagrangian quadrature to Eq. (\ref{eq:SPHPressureGrad_3}), we add the expression $p(\mathbf{x},t) \int\displaylimits_{V_x} \nabla_\mathbf{x} W_h(\mathbf{x} -  \mathbf{y}) ~\mathrm{d}\mathbf{y} = 0$, considering that the kernel gradient is anti-symmetric by definition. With that, we can state
\begin{eqnarray}
    &-&\int\displaylimits_{V_x} \nabla_\mathbf{y} p(\mathbf{y},t) W_h(\mathbf{x} -  \mathbf{y})~\mathrm{d}\mathbf{y} \label{eq:SPHPressureGrad_4} \\
    = &-&  \int\displaylimits_{V_x} ( p(\mathbf{y},t) + p(\mathbf{x},t) ) \nabla_\mathbf{x} W_h(\mathbf{x} -  \mathbf{y}) ~\mathrm{d}\mathbf{y}~. \nonumber
\end{eqnarray}
If a Lagrangian quadrature is applied to Eq. (\ref{eq:SPHPressureGrad_4}), adapting the particle notation with index $i$ \& $j$ like for the density above, one finds the following approximation for the averaged pressure gradient
\begin{equation}
    -\int\displaylimits_{V_x} \nabla_\mathbf{y} p(\mathbf{y},t) W_h(\mathbf{x} -  \mathbf{y})~\mathrm{d}\mathbf{y} 
    \approx -   \sum_{j=1}^{N_{ngb}} ( p_j + p_i ) \nabla W_{h,ij} V_j~.
    \label{eq:SPHPressureGrad_5}
\end{equation}
From this discretized form, the operation in Eq. (\ref{eq:SPHPressureGrad_4}) becomes comprehensible. It generates an anti-symmetric pressure force between particle $i$ \& $j$ in accordance with Newtons's third law \cite{Monoghan_2005}. This leads to momentum conservation in the discretized transport equations and is often derived in the SPH community by a variational principle for an ideal Euler fluid, e.g. Refs. \cite{Monoghan_2005, Springel_2010, Price_2012}. However, it is important to realize that the fluid element pressures $p_i$ \& $p_j$ in Eq. (\ref{eq:SPHPressureGrad_5}) are unknowns for the super fluid elements in the LES framework. Hence, the only option to estimate these quantities is by a replacement with the averaged pressures of the approximated super fluid elements itself, exemplary  \cite{Zhu_2015, Sigalotti_2021}
\begin{equation}
    p_i = \overline{p}_i + \mathcal{O} (h^2) + \mathcal{O} (N_{ngb}^{-\gamma})~, \quad \gamma \in [\frac{1}{2}; 1]~.
    \label{eq:SPHPressureGrad_6}
\end{equation}
This is a crucial step in order to obtain an expression fully consistent with the well-known SPH formulation, but certainly introduces two errors: The first showing a non-local $\sim h^2$ dependence due to the super fluid element replacement and the second showing a $\sim N_{ngb}^{-\gamma}$ dependence with $\gamma \in [\frac{1}{2}; 1]$ due to the Lagrangian quadrature. Summarizing, one finally obtains a well-known SPH approximation of the LES pressure gradient from the Lagrangian quadrature, which reads
\begin{equation}
    -\int\displaylimits_{V_x} \nabla_\mathbf{y} p(\mathbf{y},t) W_h(\mathbf{x} -  \mathbf{y})~\mathrm{d}\mathbf{y} 
    \approx - \sum_{j=1}^{N_{ngb}} ( \overline{p}_j + \overline{p}_i ) \nabla W_{h,ij} V_j~.
    \label{eq:SPHPressureGrad_7}
\end{equation}
and comes with the usual SPH peculiarity. The formulation depends on the pressure level $p_{ref}$ of the EOS in Eq. (\ref{eq:LES_EOS}) and breaks its gauge invariance but simultaneously introduces an implicit particle regularization based on the local particle order \cite{Price_2012, Colagrossi_2012}.

\subsection{\label{subsec:ViscousStress}The Averaged Viscous Stress Term}

For the viscous stress term in Eq. (\ref{eq:LES_Momentum}), the same manipulations can be applied in order to find an expression equivalent to Eq. (\ref{eq:SPHPressureGrad_3}). It finally reads
\begin{eqnarray}
    && \int\displaylimits_{V_x} div_\mathbf{y} [ 2\nu \rho \boldsymbol{D}](\mathbf{y}, t) W_h(\mathbf{x} -  \mathbf{y})~\mathrm{d}\mathbf{y} \label{eq:SPHViscDiv_1} \\
    = ~&& \int\displaylimits_{V_x}  [ 2\nu \rho \boldsymbol{D}](\mathbf{y}, t) \nabla_\mathbf{x} W_h(\mathbf{x} -  \mathbf{y}) ~\mathrm{d}\mathbf{y}~. \nonumber
\end{eqnarray}
We could theoretically continue as in the last section for the averaged pressure gradient, which would lead to an SPH approximation for the viscous stress term similar to the one presented in Ref. \cite{Sijacki_2006}. However, another formulation for strongly subsonic flows is much more common in the SPH community, which is conserving angular momentum in discretized form and not only in the continuum limit. Thus, we change our strategy for the viscous stress term. Therefore, it should be recognized that the tensor $2\nu \rho \boldsymbol{D}$ in Eq. (\ref{eq:SPHViscDiv_1}) depends on $\mathbf{y}$ and the integration is performed in respect to $\mathbf{y}$ as well. Hence, the $\nabla_\mathbf{x}$ operator and integration can be interchanged. Assuming $\nu=const$ and using the abbreviation defined in Eq. (\ref{eq:Filter}), it exactly yields
\begin{equation}
    \int\displaylimits_{V_x} div_\mathbf{y} [ 2\nu \rho \boldsymbol{D}](\mathbf{y}, t) W_h(\mathbf{x} -  \mathbf{y})~\mathrm{d}\mathbf{y} = 2\nu div_\mathbf{x} [ \overline{\rho \boldsymbol{D}} ] (\mathbf{x}, t)~.
    \label{eq:SPHViscDiv_2}
\end{equation}
Equation Eq. (\ref{eq:SPHViscDiv_2}) can be rearranged with the Favre average in Eq. (\ref{eq:Favre}), resulting in
\begin{equation}
    \int\displaylimits_{V_x} div_\mathbf{y} [ 2\nu \rho \boldsymbol{D}](\mathbf{y}, t) W_h(\mathbf{x} -  \mathbf{y})~\mathrm{d}\mathbf{y} = 2\nu div_\mathbf{x} [ \overline{\rho} \tilde{\boldsymbol{D}} ] (\mathbf{x}, t)~.
    \label{eq:SPHViscDiv_3}
\end{equation}
Since we are interested in strongly subsonic flows, the assumption of weak spatial changes in $\rho$ is viable. Consequently, it seems likely that the spatial changes of $\overline{\rho}$ are even weaker or negligible compared to spatial changes in $\tilde{\boldsymbol{D}}$. Thus, $div_\mathbf{x}$ and $\overline{\rho}$ can be interchanged and we assume that for the dynamic visosity $\eta := \nu \overline{\rho} = const$. Then, one arrives at \cite{Landau_1991}
\begin{equation}
    2\nu div_\mathbf{x} [ \overline{\rho} \tilde{\boldsymbol{D}} ] (\mathbf{x}, t) = 2 \eta div_\mathbf{x}[ \tilde{\boldsymbol{D}} ] (\mathbf{x}, t) = \eta \Delta \tilde{\mathbf{v}} (\mathbf{x}, t)~.
    \label{eq:SPHViscDiv_4}
\end{equation}
So far all performed manipulations are exact, given that the assumptions made are valid. In order to discretize the Laplacian in Eq. (\ref{eq:SPHViscDiv_4}), a technique from the SPH community is used, which was firstly introduced by Brookshaw and reduces the sensitivity of the discretization to the local particle order. The main idea is to approximate second order derivatives by a non-local integral expression \cite{Brookshaw_1985}. It was further developed by Espa\~nol \& Revenga \cite{Espanol_2003} and Hu \& Adams \cite{Hu_2006_Letter} and results in the following estimate for the Laplacian of the Favre averaged velocity field \cite{Colagrossi_2017}
\begin{eqnarray}
    \int\displaylimits_{V_x} \hspace{-0.2cm} &&div_\mathbf{y} [ 2\nu \rho \boldsymbol{D}](\mathbf{y}, t) W_h(\mathbf{x} -  \mathbf{y})~\mathrm{d}\mathbf{y} = \eta \Delta \tilde{\mathbf{v}} (\mathbf{x}, t) \label{eq:SPHViscDiv_5} \\
    &=& ~ 2(2+n) \eta \int\displaylimits_{V_x}  \frac{(\tilde{\mathbf{v}} (\mathbf{x}, t) - \tilde{\mathbf{v}} (\mathbf{y}, t)) \cdot (\mathbf{x} -  \mathbf{y})}{(\mathbf{x} - \mathbf{y})^2} \nabla_\mathbf{x} W_h(\mathbf{x} -  \mathbf{y}) ~\mathrm{d}\mathbf{y} \nonumber \\
    &+& \mathcal{O} (h^2) ~, \nonumber
\end{eqnarray}
with $n$ as dimension of the problem. Based on Eq. (\ref{eq:SPHViscDiv_5}), finally a Lagrangian quadrature for a finite number of approximation particles can be applied. With particle index notation one finds an estimate of Eq. (\ref{eq:SPHViscDiv_5}), which should formally be exact for $N_{ngb} \to \infty$, and conserves angular momentum also on the discrete level, as the resulting inter-particle forces are collinear to particle interaction lines \cite{Colagrossi_2017}. It gives the well-known SPH expression
\begin{eqnarray}
    \int\displaylimits_{V_x} \hspace{-0.2cm} &&div_\mathbf{y} [ 2\nu \rho \boldsymbol{D}](\mathbf{y}, t) W_h(\mathbf{x} -  \mathbf{y})~\mathrm{d}\mathbf{y}  \label{eq:SPHViscDiv_6} \\
    &\approx& ~ 2(2+n) \eta  \sum_{j=1}^{N_{ngb}}  \frac{(\tilde{\mathbf{v}}_i - \tilde{\mathbf{v}}_j ) \cdot (\mathbf{x}_i -  \mathbf{y}_j)}{(\mathbf{x}_i - \mathbf{y}_j)^2} \nabla W_{h,ij} V_j ~. \nonumber 
\end{eqnarray}

\subsection{\label{subsec:SFS}The Subfilter Stress Term}

In order to discretize the subfilter stress term $div_{\mathbf{x}} \left[ \boldsymbol{\tau}_{SFS} \right](\mathbf{x}, t)$ in Eq. (\ref{eq:LES_Momentum}), we start by replacing this term by its averaged, non-local counterpart \cite{Zhu_2015, Sigalotti_2021}
\begin{equation}
    div_{\mathbf{x}} \left[ \boldsymbol{\tau}_{SFS} \right](\mathbf{x}, t) = \int\displaylimits_{V_x} div_{\mathbf{y}} \left[ \boldsymbol{\tau}_{SFS} \right](\mathbf{y}, t) W_h(\mathbf{x} -  \mathbf{y})~\mathrm{d}\mathbf{y} + \mathcal{O}(h^2)~.
    \label{eq:SPH_SFS_1}
\end{equation}
This opens the opportunity to shift the effect of the $div_{\mathbf{x}}$ operator to the kernel itself, which can be computed analytically. By using the same arguments as for the averaged pressure gradient in Eq. (\ref{eq:SPHPressureGrad_1}) \& Eq. (\ref{eq:SPHPressureGrad_2}), one finds the subfilter stress term counterpart of Eq. (\ref{eq:SPHPressureGrad_3}). It reads
\begin{equation}
    div_{\mathbf{x}} \left[ \boldsymbol{\tau}_{SFS} \right](\mathbf{x}, t) \approx \int\displaylimits_{V_x}  \boldsymbol{\tau}_{SFS} (\mathbf{y}, t) \nabla_{\mathbf{x}} W_h(\mathbf{x} -  \mathbf{y})~\mathrm{d}\mathbf{y}~.
    \label{eq:SPH_SFS_2}
\end{equation}
Analogously to Eq. (\ref{eq:SPHPressureGrad_4}), we can add $\boldsymbol{\tau}_{SFS}(\mathbf{x},t) \int\displaylimits_{V_x} \nabla_\mathbf{x} W_h(\mathbf{x} -  \mathbf{y}) ~\mathrm{d}\mathbf{y} = 0$ in Eq. (\ref{eq:SPH_SFS_2}), to finally create an anti-symmetric force according to Newton's third law after the Lagrangian quadrature. This results in
\begin{equation}
    div_{\mathbf{x}} \left[ \boldsymbol{\tau}_{SFS} \right](\mathbf{x}, t) \approx \int\displaylimits_{V_x} ( \boldsymbol{\tau}_{SFS} (\mathbf{y}, t) + \boldsymbol{\tau}_{SFS} (\mathbf{x}, t) ) \nabla_{\mathbf{x}} W_h(\mathbf{x} -  \mathbf{y})~\mathrm{d}\mathbf{y}~.
    \label{eq:SPH_SFS_3}
\end{equation}
Applying a Lagrangian quadrature to Eq. (\ref{eq:SPH_SFS_3}) and using particle index notation, the approximation takes the form
\begin{equation}
    div_{\mathbf{x}} \left[ \boldsymbol{\tau}_{SFS} \right](\mathbf{x}, t) \approx \sum_{j=1}^{N_{ngb}} ( \boldsymbol{\tau}_{SFS,j} + \boldsymbol{\tau}_{SFS,i} ) \nabla W_{h,ij}~V_j~.
    \label{eq:SPH_SFS_4}
\end{equation}
However, as usual for LES, the most interesting part of the subfilter stress term approximation consists in finding an estimate for the SFS tensor $\boldsymbol{\tau}_{SFS}$ itself. 

The standard option is to employ the eddy viscosity concept in connection with Boussinesq’s hypothesis \cite{Sagaut_2006, Schmitt_2007, Silvis_2017, Moser_2021}. Although this class of SFS models is known to oversimplify physical effects below the subfilter scale \cite{Sagaut_2006, Schmitt_2007, Silvis_2017, Moser_2021}, it is compliant with the dissipative statistical property of the energy cascade \cite{Eyink_2018, Moser_2021}. This means that kinetic energy is mostly transferred from larger to smaller scales. As our goal is to eliminate numerical noise in favor of the large scales, explained in Sec. \ref{sec:SFS}, we deem eddy viscosity models as appropriate for this study. Therefore, the SFS tensor can be approximately expressed as \cite{Garnier_2009}
\begin{equation}
    \boldsymbol{\tau}_{SFS}(\mathbf{x}, t) \approx - 2 \nu_t \overline{\rho}  \tilde{\boldsymbol{D}}(\mathbf{x}, t)~,
    \label{eq:SPH_SFS_7}
\end{equation}
assuming that the isotropic part of the tensor is negligible for strongly subsonic flows. In Eq. (\ref{eq:SPH_SFS_7}), the scalar field $\nu_t$ denotes the eddy viscosity and the tensor field $\tilde{\boldsymbol{D}}$ the Favre averaged strain rate. The latter is defined by
\begin{equation}
    \tilde{\boldsymbol{D}}(\mathbf{x}, t) := \frac{1}{2}( \tilde{\boldsymbol{J}} + \tilde{\boldsymbol{J}}^T )(\mathbf{x}, t)~
    \label{eq:SPH_SFS_8}
\end{equation}
with $\tilde{\boldsymbol{J}}$ representing the Favre averaged velocity field Jacobian. With the aid of the techniques described above, one can construct a Lagrangian quadrature approximation for the Jacobian, which is well-known in the SPH community and first order consistent in the continuum limit \cite{Price_2012}, namely for a specific particle
\begin{equation}
    \tilde{\boldsymbol{J}}_i := \tilde{\boldsymbol{J}}(\mathbf{x}_i, t) \approx \sum_{j=1}^{N_{ngb}} (\tilde{\mathbf{v}}_j - \tilde{\mathbf{v}}_i) \nabla W_{h,ij}^T V_j~.
    \label{eq:SPH_SFS_9}
\end{equation}
Hence, Eq. (\ref{eq:SPH_SFS_8}) is defined on the particle level as well. 

For the remaining unknown field $\nu_t$ in Eq. (\ref{eq:SPH_SFS_7}), we will explore two different models. The first model will be the standard Smagorinsky model with constant $C_S=0.15$ and the filter width $\Delta$ given by \cite{Sagaut_2006, Garnier_2009, Silvis_2017}
\begin{equation}
    \nu_t := (C_S \Delta)^2 \sqrt{2~\mathrm{tr}\{ \tilde{\boldsymbol{D}}^2 \}}~,
    \label{eq:SPH_SFS_10}
\end{equation}
in which $\tilde{\boldsymbol{D}}$ can be computed from Eqs. (\ref{eq:SPH_SFS_8}) \& (\ref{eq:SPH_SFS_9}) and $\mathrm{tr}\{\cdot \}$ denotes the trace operation. The second model considered, is the $\sigma$-model developed by Nicoud \emph{et al.} \cite{Nicoud_2011}. It overcomes some severe drawbacks of the Smagorinsky model, e.g. it guarantees vanishing subfilter dissipation in laminar regions and proper wall scaling \cite{Nicoud_2011, Silvis_2017}. Based on the singular values $\sigma_k$, $k \in \{1,2,3\}$, of the tensor $\tilde{\boldsymbol{J}}^T \tilde{\boldsymbol{J}}$, the alternative eddy viscosity model with the model constant $C_\sigma = 1.35$ reads \cite{Nicoud_2011, Silvis_2017}
\begin{equation}
    \nu_t := (C_\sigma \Delta)^2 \frac{\sigma_3 (\sigma_1 - \sigma_2) (\sigma_2 - \sigma_3)}{\sigma_1^2}~.
    \label{eq:SPH_SFS_11}
\end{equation}

It should be noted that the modelled subfilter stress term according to Eq. (\ref{eq:SPH_SFS_4}) is only angular momentum conserving in the continuum limit, but not on the discrete particle level \cite{Sijacki_2006}. Therefore, for comparative reasons, we will additionally consider a heuristic augmentation of the averaged viscous stress term in Eq. (\ref{eq:SPHViscDiv_6}) to variable eddy viscosity based on the ideas of Ref.\cite{Cleary_1999}. Exemplary, it is utilized in the SPH-LES works of Di Mascio \emph{et al.} \cite{DiMascio_2017} and Antuono \emph{et al.} \cite{Antuono_2021}. The inherently angular momentum conserving alternative of Eq. (\ref{eq:SPH_SFS_4}) is
\begin{eqnarray}
    div_{\mathbf{x}} \left[ \boldsymbol{\tau}_{SFS} \right](\mathbf{x}, t) &\approx&  \label{eq:SPH_SFS_12} \\
    2(2+n)  \sum_{j=1}^{N_{ngb}} &\overline{\rho}_i& \overline{\rho}_j \frac{ \nu_{t,i} + \nu_{t,j} }{ \overline{\rho}_{i} + \overline{\rho}_{j} }  \frac{(\tilde{\mathbf{v}}_i - \tilde{\mathbf{v}}_j ) \cdot (\mathbf{x}_i -  \mathbf{y}_j)}{(\mathbf{x}_i - \mathbf{y}_j)^2} \nabla W_{h,ij} V_j ~. \nonumber
\end{eqnarray}
and will be called Monoghan-Cleary-Gingold (MCG) form according to Ref. \cite{Colagrossi_2017}.

\section{\label{sec:AppendixB}\textcolor{black}{Influence of Reynolds Number}}

\textcolor{black}{Apart from the conclusions presented in the main part of our work, one can further question how the viscosity or $Re$ number influences the shape of the energy spectra.}
\begin{figure}[h]
\includegraphics[width=3.in, clip, trim= 0cm 0.9cm 0cm 0cm]{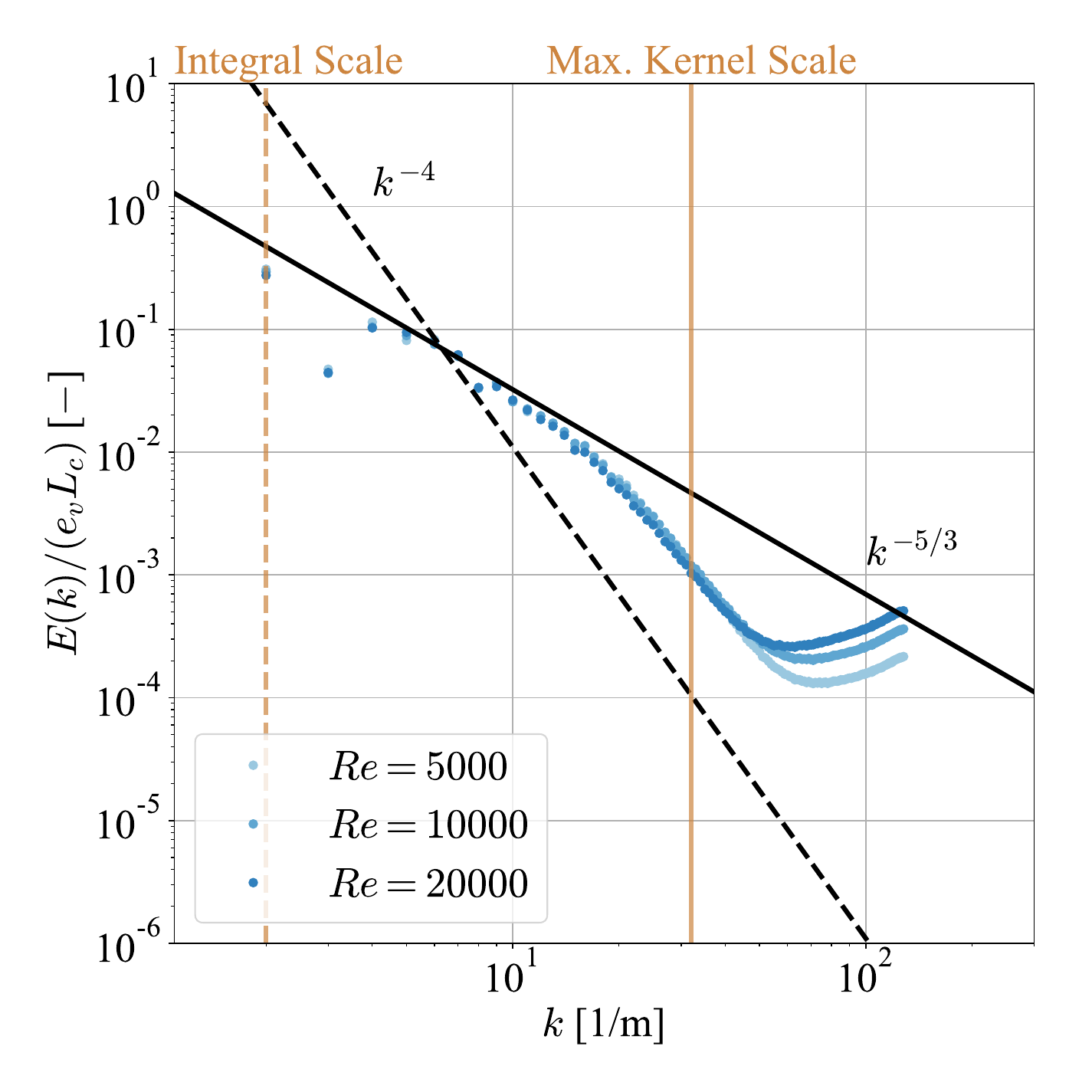}
\caption{\textcolor{black}{Kinetic energy spectra at $t=14~\mathrm{s}$ for different $Re$.}}
\label{fig:B1_Reynolds}
\end{figure}

\textcolor{black}{From a physical perspective, it is well known that the width of the inertial range scales with $\sim Re^{3/4}$, see Ref. \cite{Bailly_2015}. However, our study focuses on strongly underresolved Lagrangian SPH-LES simulations were viscous forces are not correctly represented due to a lack of resolution. In order to restore viscous dissipation characteristics using a sixth order finite difference method, as in the study of Dairay et al. \cite{Dairay_2017} for $Re=10^4$, a resolution of $2048^3$ elements was required. Contrary, our results demonstrate that the SPH inertial ranges are controlled by numerical dissipation and dispersion effects representing an implicit SFS model. Therefore, we do not expect significant change with different $Re$ as long as the kernel scale lies within the inertial range. This can be mathematically proved using the Chauchy-Schwartz inequality \cite{Eyink_2018} and is confirmed in FIG. \ref{fig:B1_Reynolds} for Case 3 in TABLE \ref{tab:Cases} with halved and doubled $Re$. For the given resolution the inertial range scaling prevails in the same wavenumber range. However, it is interesting to note that higher viscosities are beneficial in terms of the energy levels of the artificial thermalization. As a consequence, scales larger than the kernel relatively contain slightly more energy at lower $Re$, although this effect is evidently not very pronounced.}

\section{\label{sec:AppendixC}\textcolor{black}{Influence of Kernel Size}}

\begin{figure}[h]
\includegraphics[width=3.in, clip, trim= 0cm 0.9cm 0cm 0cm]{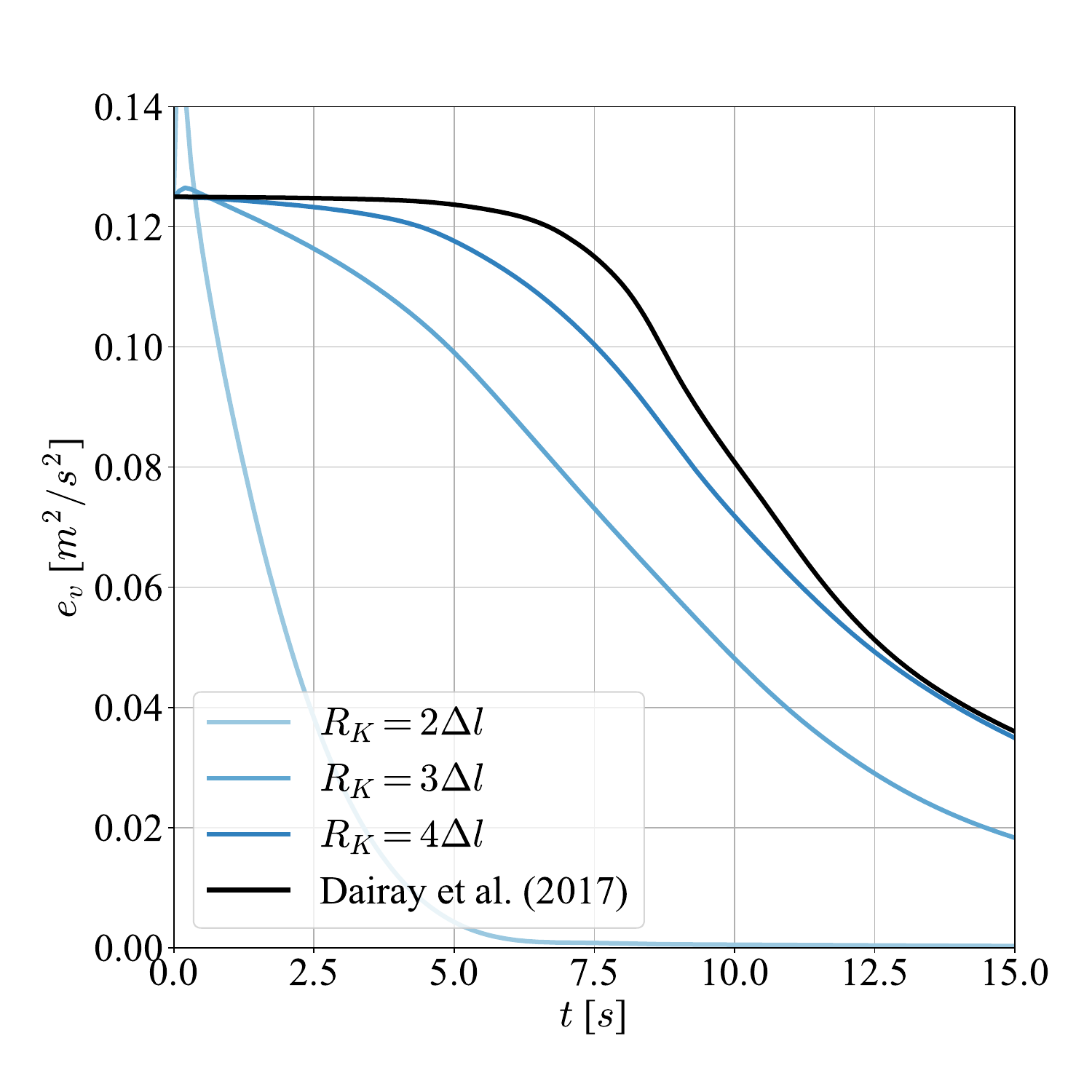}
\caption{\textcolor{black}{Temporal evolution of the density weighted averaged kinetic energy $e_v$ for different $R_K$ with $N=256^3$.}}
\label{fig:C1_RK_Viscosity_Energy}
\end{figure}

\textcolor{black}{In the main part of our work we only discuss cases with a kernel diameter of $D_K=8\Delta l$ or radius of $R_K=4\Delta l$. This is motivated by the observation that adequate numerical convergence can hardly be observed for the given problem with smaller values, highlighting the paramount importance of $N_{ngb}$. To give an idea, the temporal evolution of the density weighted averaged kinetic energy $e_v$ is depicted in FIG. \ref{fig:C1_RK_Viscosity_Energy} for smaller $R_K$ but otherwise same conditions as in Case 3 in TABLE \ref{tab:Cases}. }

\begin{figure}[h]
\includegraphics[width=3.in, clip, trim= 0cm 0.9cm 0cm 0cm]{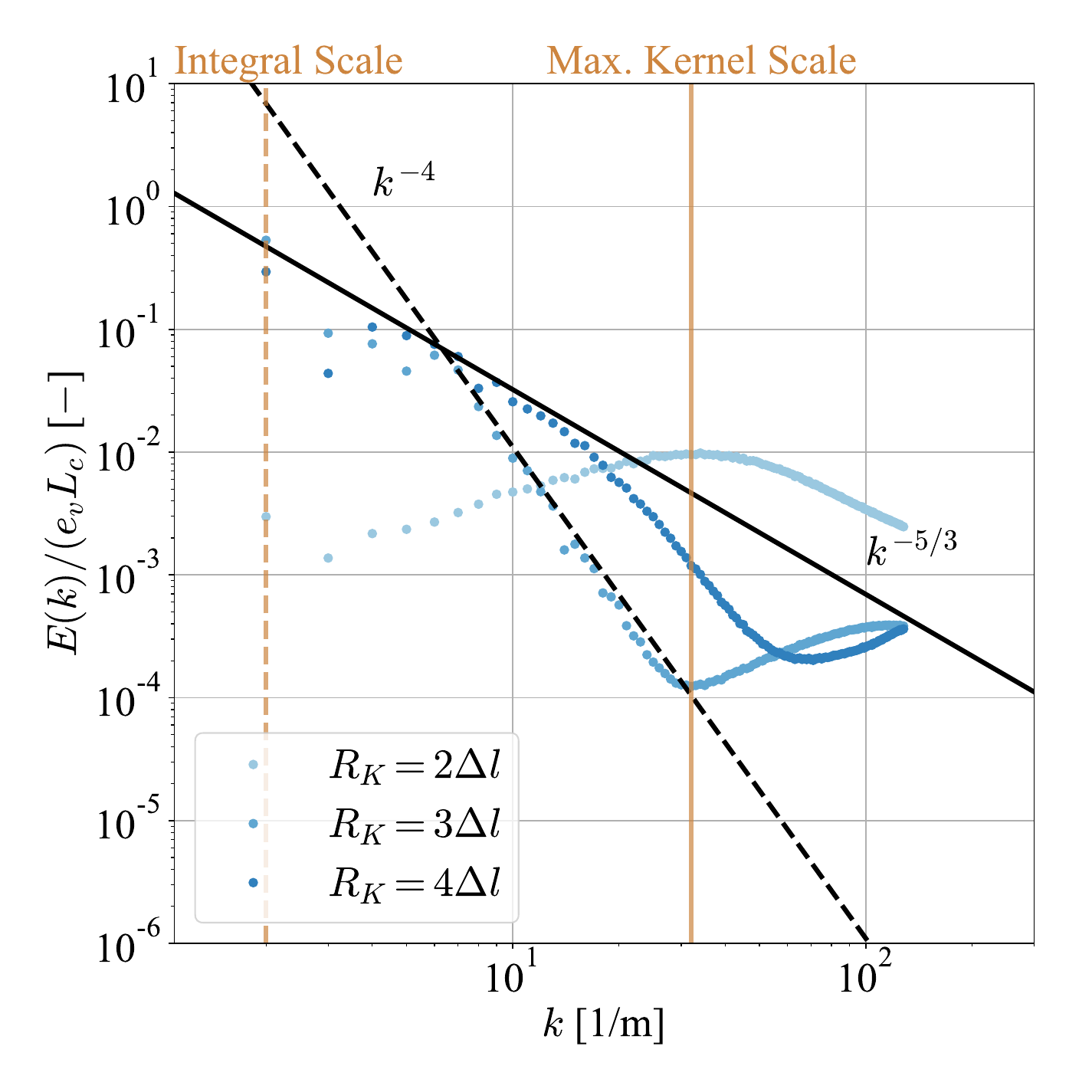}
\caption{\textcolor{black}{Kinetic energy spectra at $t=14~\mathrm{s}$ for different $R_K$ with $N=256^3$.}}
\label{fig:C1_RK_Viscosity_Spectrum}
\end{figure}

\textcolor{black}{Apparently, for a given $N$, the dissipation rate is strongly reduced with increasing $N_{ngb}$. Likewise are the artificial compressibilty effects at the beginning of the simulation, which lead to a production of kinetic energy. In spectral space the effect is similarly pronounced, as shown in FIG. \ref{fig:C1_RK_Viscosity_Spectrum}. Only the $R_K=4\Delta l$ case shows a pronounced inertial range scaling for $N=256^3$ particles. These observations are compliant with the numerical convergence behaviour of SPH presented by Zhu \emph{et al.} \cite{Zhu_2015}. }

\bibliography{SPH_Revised_Okraschevski}

\end{document}